\documentclass[12pt]{article} 

\usepackage[colorlinks=true]{hyperref}
\usepackage[margin=1.in]{geometry}
\usepackage{graphicx}
\usepackage{amsmath}
\usepackage{amssymb}
\usepackage{booktabs}
\usepackage{caption}
\usepackage{subcaption}
\usepackage{verbatim}
\usepackage{float} 

\newcommand{\vc}[1]{{\pmb{#1}}}
\newcommand{\ra}[1]{\renewcommand{\arraystretch}{#1}}
\newcommand{\ip}[2]{\langle{#1},{#2}\rangle}
\DeclareMathOperator{\diag}{diag}

\title{Benchmarking Quantum Hardware for Training of \\Fully Visible
  Boltzmann Machines}

\author{Dmytro Korenkevych$^1$, Yanbo Xue$^2$, Zhengbing Bian$^2$, Fabian Chudak$^2$, \\ William
  G. Macready$^2$, Jason Rolfe$^2$, Evgeny Andriyash$^2$ \\ \\
$^1$ KindredAI, Vancouver, BC, Canada\\
$^2$ D-Wave Systems, Burnaby, BC, Canada}

\date{\today}

\begin{document}

\maketitle

\begin{abstract}
  Quantum annealing (QA) is a hardware-based heuristic optimization
  and sampling method applicable to discrete undirected graphical
  models. While similar to simulated annealing, QA relies on quantum,
  rather than thermal, effects to explore complex search spaces. For
  many classes of problems, QA is known to offer computational
  advantages over simulated annealing. Here we report on the ability
  of recent QA hardware to accelerate training of fully visible
  Boltzmann machines. We characterize the sampling distribution of QA
  hardware, and show that in many cases, the quantum distributions
  differ significantly from classical Boltzmann distributions.  In
  spite of this difference, training (which seeks to match data and
  model statistics) using standard classical gradient updates is still
  effective.  We investigate the use of QA for seeding Markov chains
  as an alternative to contrastive divergence (CD) and persistent
  contrastive divergence (PCD). Using $k=50$ Gibbs steps, we show that
  for problems with high-energy barriers between modes, QA-based seeds
  can improve upon chains with CD and PCD initializations. For these
  hard problems, QA gradient estimates are more accurate, and allow
  for faster learning.  Furthermore, and interestingly, even the case
  of raw QA samples (that is, $k=0$) achieved similar improvements.
  We argue that this relates to the fact that we are training a
  quantum rather than classical Boltzmann distribution in this case.
  The learned parameters give rise to hardware QA distributions
  closely approximating classical Boltzmann distributions that are
  hard to train with CD/PCD.

\end{abstract}

\section{Introduction}
\label{introduction}

In the early 1980s, a number of authors suggested that certain
computations might be accelerated with computers making use of quantum
resources
\cite{benioff80,deutsch85:_quant_theor_churc_turin_princ}. Feynman's
1981 proposal \cite{feynman82:_simul_physic_comput} suggested that
quantum systems themselves might be more efficiently modelled with
quantum computers. Over a decade later, Peter Shor devised a
polynomial-time quantum method for factoring large integers. Despite
this theoretical promise, progress towards experimental quantum
computing platforms remained limited. It was not until 1998, with the
introduction of quantum annealing (QA) \cite{kadowaki98:_quant_ising},
that a path to scalable quantum hardware emerged. While existing QA
machines are not computationally universal, QA machines are available
now at large scales and offer significant speedups for certain problem
classes \cite{v.15:_what_is_comput_value_finit_range_tunnel}. Here, we
explore the potential of QA to accelerate training of probabilistic
models.

The QA heuristic operates in a manner analogous to simulated annealing
(SA), but relies on quantum, rather than thermal, fluctuations to
foster exploration through a search space. Just as thermal
fluctuations are annealed in SA, quantum fluctuations are annealed in
QA.

With the exception of \cite{Adachi2015,Benedetti2016}, most
applications run on QA hardware have used the optimization potential
of quantum annealing.  In \cite{Adachi2015}, the focus is on training
a 4-layer deep belief network. Pre-training of each layer uses
restricted Boltzmann machines (RBMs) trained using QA via a complete
bipartite graph embedding.  \cite{Adachi2015} tested their approach
against 1-step contrastive divergence (CD) samples on a coarse-version
of MNIST and concluded that QA sped up training significantly.  In
\cite{Benedetti2016}, the authors consider training a fully connected
Boltzmann machine (BM) using QA via a complete graph embedding on the
hardware graph. They report a training speed-up compared to training
with simulated annealing directly on the complete graph. These studies
assume that the quantum hardware produces a classical Boltzmann
distribution. In contrast, in this paper we do not assume the QA
samples are Boltzmann. We demonstrate the differences between
classical Boltzmann and QA hardware samples, and explore the impact of
these differences in training fully-visible BMs in small density
estimation tasks.  Training of BMs is a natural application domain
because available QA hardware realizes Boltzmann-like distributions,
inference in BMs is known to be very hard \cite{icml2010_LongS10}, and
BMs are a building block of many generative probabilistic models
\cite{SalHinton07}.

We begin with background on QA on the annealing-based quantum system,
highlighting its practical constraints. We characterize the sampling
done by the hardware, which in some cases is Boltzmann and in other
cases differs significantly from Boltzmann. We then describe the
challenge of learning probabilistic models with BMs,
and how QA might accelerate such training. We provide benchmark
results on the learning of multimodal distributions, and quantify the
benefits that QA can offer. Lastly, we show the impact of the
non-Boltzmann nature of the D-Wave system, and how this impacts
learning.  We conclude with directions for future work.

\section{Quantum Annealing}
\label{sec:quantum-annealing}

\begin{figure}
  \centering
\begin{subfigure}[b]{0.6\textwidth}
\centering
  \includegraphics[width=0.8\textwidth]{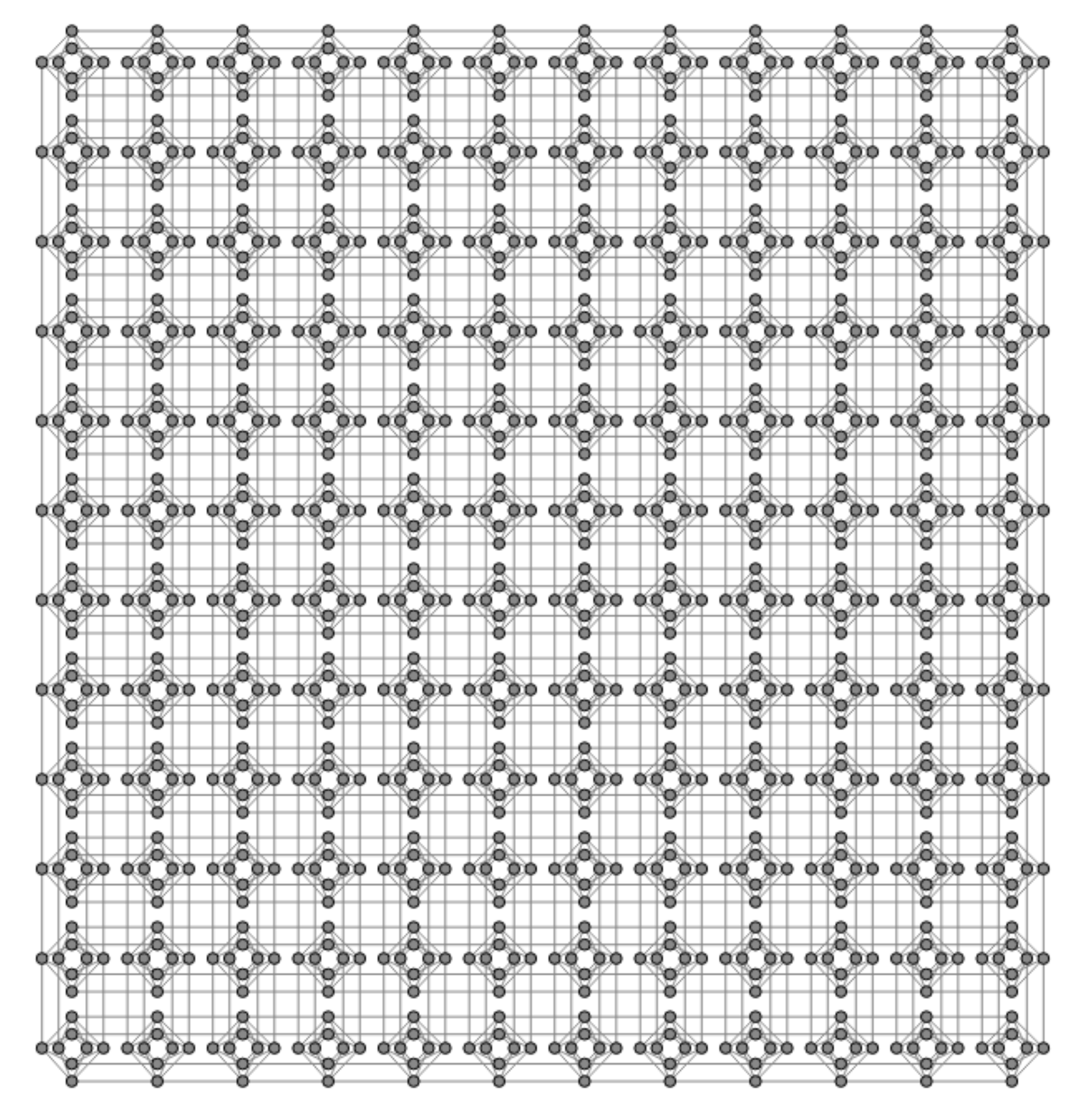}
  \caption{The $C_{12}$ Chimera graph consisting of a $12\times 12$
    array of $K_{4,4}$ bipartite unit cells. Nodes represent problem
    variables with programmable weights $h$, and edges have a
    programmable $J$ connection.}
  \label{fig:chimera}
\end{subfigure}
~
\begin{subfigure}[b]{0.35\textwidth}
  \centering \ra{1.3}
  \begin{tabular}{@{}lr@{}}
    \toprule
    programming time & 25 ms \\
    anneal time & $>5$ $\mu$s/sample \\
    readout time & 260 $\mu$s/sample \\
    \bottomrule
  \end{tabular}
  \caption{Typical timing data of the 2000-qubit D-Wave system.}
  \label{tab:timing}
\end{subfigure}
\caption{2000-qubit D-Wave system parameters.}
\end{figure}

QA uses quantum-mechanical processes to minimize and sample from
energy-based models. The D-Wave machine implements the Ising model
energy function:\footnote{Vectors are indicated in lowercase
  bold font, and matrices in uppercase bold font.}
\begin{equation*}
  E(\vc{s}) = \sum_{v\in \mathcal{V}} h_v s_v + \sum_{(v_1,v_2)\in \mathcal{E}} J_{v_1,v_2}
  s_{v_1} s_{v_2} \quad \text{with $s_v\in\{-1,+1\}$}
\end{equation*}
with variable connectivity defined by a graph
$G=(\mathcal{V},\mathcal{E})$. The 2000-qubit D-Wave system allows for up
to $|\mathcal{V}|=1152$ variables with sparse bipartite connectivity.
The connectivity graph of the D-Wave device is called Chimera, and
denoted $C_n$. $C_n$ consists of an $n\times n$ array of $K_{4,4}$
unit cells with connection between unit cells as in
Fig.~\ref{fig:chimera}, which shows a $C_{12}$ graph.
The tree-width of $C_{12}$ graph is 48, so exact inference is practically
impossible. It is simple to convert $\pm 1$ valued spins $s_v$ to
Boolean-valued variables $x_v = (1+s_v)/2$ so that $E(\vc{s})$ also
defines a BM with energy $E(\vc{x})$ and the same sparse bipartite
connectivity.

Quantum mechanics replaces the energy function with a linear operator
acting on states $\vc{s}$ and returning new states $\vc{s}'$. This
energy operator is described by the Hamiltonian, a
$2^{|\mathcal{V}|}\times 2^{|\mathcal{V}|}$ matrix $\vc{H}$ whose components are indexed
by $(\vc{s},\vc{s}')$. The diagonal elements of $\vc{H}$ record the
energy of the corresponding states, \textit{i.e.},
$H_{\vc{s},\vc{s}} = E(\vc{s})$, and the off-diagonal elements of
$\vc{H}$ act to transform states. In the D-Wave machine the only
allowed off-diagonal contributions are those which flip bits,
\textit{i.e.} for $\vc{s}\neq \vc{s}'$
\begin{equation*}
H_{\vc{s},\vc{s}'}=\begin{cases} \Delta & \text{if $\vc{s}$ and
    $\vc{s}'$ differ in one bit} \\ 0 & \text{otherwise} \end{cases}.
\end{equation*}
Quantum processes favor states corresponding to the eigenvectors of
low-energy eigenvalues of $\vc{H}$. Thus, at zero temperature when
$\Delta = 0$, quantum evolution corresponds to uniform sampling within
the eigenspace corresponding to the lowest eigenvalue (energy) of
$\vc{H}$. However, $\Delta\neq 0$ gives rise to eigenvectors that are
linear combinations of basis vectors. These states are called
superpositions, and are interpreted as follows. An arbitrary
superposition is written as
$\vc{v} \equiv \sum_{\vc{s}} a_\vc{s} \vc{e}_{\vc{s}}$ where
$a_{\vc{s}}$ is a weight (often called an amplitude), and
$\vc{e}_\vc{s}$ is the basis vector corresponding to state
$\vc{s}$. In superposition $\vc{v}$ any particular state, $\vc{s}$, is
observed with probability proportional to $|a_\vc{s}|^2$. Thus, the
quantum state $\vc{v}$ implicitly encodes $O(2^{|\mathcal{V}|})$ degrees of
freedom. Superposition states are unavailable in non-quantum devices,
and are a source of the speedups seen in quantum computations. In
hardware like the D-Wave annealer, superposition states are generated
by physical processes and do not need to be simulated.

In QA algorithms, $\vc{H}$ is varied over time so that\footnote{$[p]$
  is Iverson's bracket defined to be 1 if predicate $p$ is true, and 0
  otherwise.}
\begin{equation}
H_{\vc{s},\vc{s}'}(t) = \mathcal{A}(t/\tau) \Delta [\text{$\vc{s}$ and $\vc{s}'$ differ in one bit}] + \mathcal{B}(t/\tau) E(\vc{s}) [\vc{s}=\vc{s}'].
\label{Ht}
\end{equation}
The time-dependent weightings $\mathcal{A}/\mathcal{B}$ are monotonically
decreasing/increasing and satisfy $\mathcal{A}(1)=0$ and $\mathcal{B}(0)=0$, so that we
evolve from $\vc{H}(0)$ --- which has no diagonal energy contribution, and
which assigns equal probability to all states
($|a_{\vc{s}}| = 1/\sqrt{2^{|\mathcal{V}|}}$) --- to the Ising energy function
$\vc{H}(\tau) \propto \diag\bigl(E(\vc{s})\bigr)$. The decreasing quantum
effects mediated by $\mathcal{A}$ give rise to the name quantum annealing. For
certain classes of optimization problems, quantum annealing can be
dramatically faster than simulated annealing
\cite{j.15:_bench_quant_anneal_proces_time_target_metric,v.15:_what_is_comput_value_finit_range_tunnel}.

On the 2000-qubit D-Wave system, the annealing time $\tau$ is programmable (the
default anneal time is 20 $\mu$s). A single sample is then measured
(drawn) at time $\tau$, and the process is repeated in an i.i.d.\ fashion
for subsequent anneals. On the first anneal, the parameters $\vc{h}$
and $\vc{J}$ must be specified, requiring a programming time around
$25$ ms. Further timing data of the 2000-qubit D-Wave system are listed in
Fig.~\ref{tab:timing}.

The Ising Hamiltonian described above is a zero temperature
($\beta=\infty$) idealization of real-world complexities. Important
deviations from ideality arise from:
\begin{itemize}
\item \textit{Finite temperature:} QA hardware does not operate at
  zero temperature. In units where the parameters lie in the interval
  $-1\le h_v\le 1$ and $-1\le J_{v,v'}\le 1$, the effective hardware
  temperature $T_{\text{HW}}$ is problem dependent and usually between $1/5$ and
  $1$.\footnote{The sampling distribution is not Boltzmann so the
    notion of temperature as it appears in a Boltzmann distribution is
    ill-defined, and many factors beyond physical temperature
    contribute to an effective ``temperature.''}
\item \textit{Parameter misspecification:} During programming the
  $\vc{h}/\vc{J}$ parameters are subject to additive Gaussian noise
  having standard deviations $\sigma_h\approx 0.03$ and
  $\sigma_J\approx 0.025$ respectively. Additionally, in spite of
  calibration of the device, small systematic deviations from the
  idealized Ising model arise because the Ising model is only an
  approximation to the true low-energy physics.
\item \textit{Dynamics:} The quantum mechanical evolution of the
  annealing process cannot be simulated at large scales (even for
  idealized models), and quantum effects can cause significant
  deviations from the classical Boltzmann distribution. A better
  approximation is obtained using the density matrix of the quantum
  Boltzmann distribution $\vc{\rho} = \exp(-\beta\vc{H})/Z(\beta)$,
  but even this approximation fails to capture the out-of-equilibrium
  effects of rapid annealing within the D-Wave device \cite{Raymond2016}.
\end{itemize}
In spite of these complexities, it remains true that QA hardware
rapidly produces i.i.d.\ low energy samples from programmable
Chimera-structured energy models. Here, we explore whether this
capability can be harnessed for efficient learning of
Chimera-structured BMs. As our interest is on the
sampling aspects of learning, we focus on fully visible models
to avoid the confounding influence of multimodal likelihood functions.

\section{QA Versus Boltzmann Sampling}
\label{sec:sampl-distr}

To begin, we explore the QA sampling distributions. As a rough
characterization, we might expect a Boltzmann distribution
$B(\vc{s})=\exp\bigl( -\beta E(\vc{s}) \bigr)/Z(\beta)$, and indeed
for some problems this is a good description. However, the Boltzmann
distribution assumes classical statistics, and numerous experiments
have confirmed the quantum nature of the D-Wave systems
\cite{t.14:_entan,v.15:_what_is_comput_value_finit_range_tunnel}. With
different choices of energy functions we can clearly expose its
quantum properties.

\begin{figure}
\centering
\begin{subfigure}[b]{0.3\textwidth}
\includegraphics[width=\textwidth]{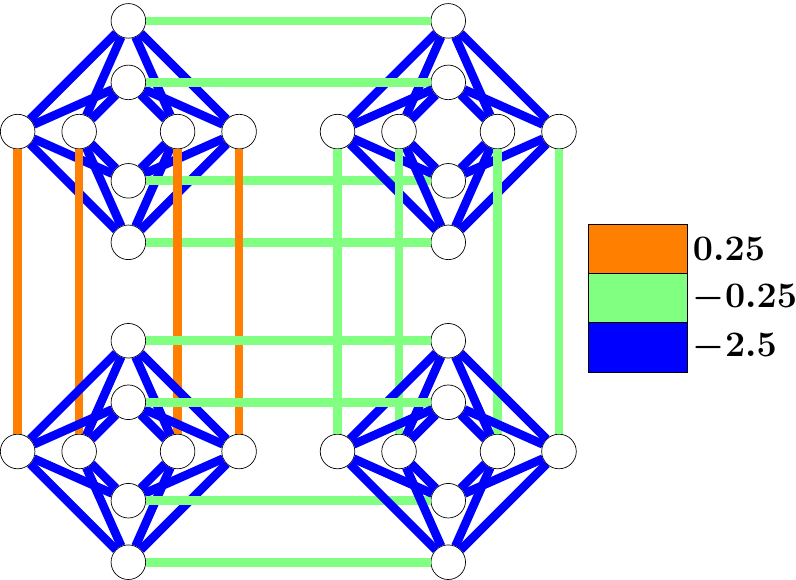}
\caption{FCL-1 problem. Edge colors represent weights of
    $J$ connections, and all $h$ biases are 0.}
\label{fig:flc1}
\end{subfigure}
~
\begin{subfigure}[b]{0.65\textwidth}
\includegraphics[width=\textwidth]{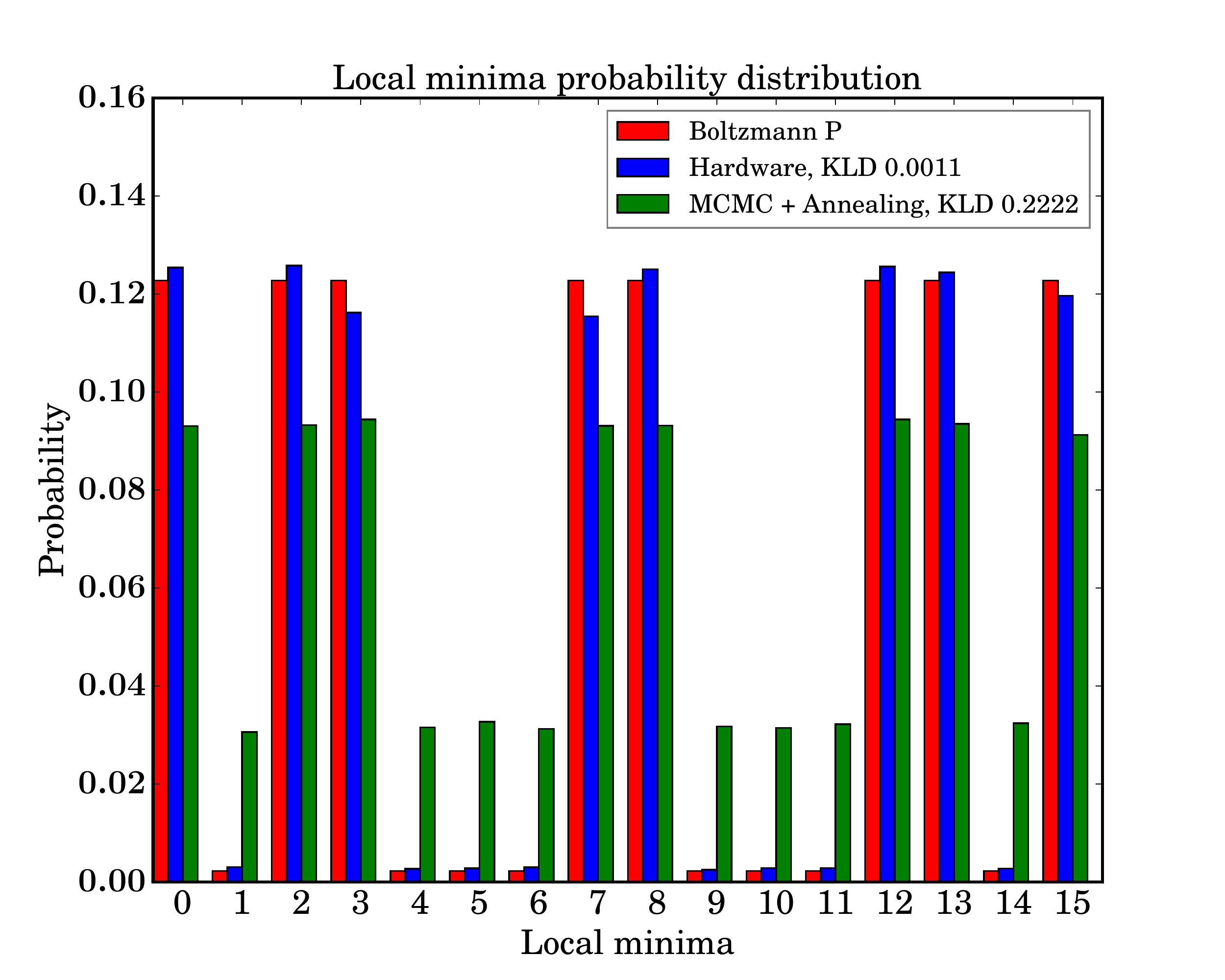}
\caption{Probabilities of local optima for $10^5$ QA and MCMC
  annealing samples. Red bars represent exact Boltzmann probabilities
  of local optima, while blue and green bars represent empirical
  probabilities of QA and MCMC samples.}
\label{fig:sampling_flc1}
\end{subfigure}
\caption{FCL-1.}
\end{figure}

Consider the Ising model illustrated on Fig.~\ref{fig:flc1}. The model
consists of 4 unit cells. The variables within each unit cell are
strongly ferromagnetically\footnote{Ferromagnetic
  ($J_{\text{intra}}<0$) connections induce neighbouring spins to take
  the same value in low energy states.} connected with connection
weights of~$J_{\text{intra}}=-2.5$. The connections between unit cells
form a frustrated loop, and have weights $J_{\text{inter}}$ 10 times
weaker in magnitude than the intra-cell connections. The $h$ weights
on all variables are zero. We call this a frustrated loop of clusters
problem, and reference it as FCL-1.

The energy landscape of FCL-1 has 16 local minima corresponding to the
$2^4$ possible configurations of four unit cells (each variable within
a unit cell takes the same value in low-energy states). These 16 local
minima are separated by high-energy barriers. To cross a barrier, an
entire unit cell must be flipped, incurring an energy penalty of
$16J_{\text{intra}}$. Among the 16 local minima, 8 are ground states
and 8 are excited states, and the energy gap between ground and
excited states is 4.

The energy barriers make it very difficult for any single-spin-flip
Markov chain Monte Carlo (MCMC) algorithm to move between valleys. To
draw approximate Boltzmann samples from FCL-1, we ran $10^5$ MCMC
chains from random initializations, and updated each using blocked
Gibbs sampling with $10^4$ Gibbs updates, annealed over 1000
temperature steps.\footnote{Blocked Gibbs sampling without annealing
  performed much more poorly.} The inverse temperature steps were set
uniformly over the interval $\beta = [0.01, 1.0]$ so there are 10
blocked Gibbs updates at each $\beta$.

Under FCL-1, we also generated $10^5$ QA samples\footnote{We used 100
  random spin-reversal transformations as suggested by D-Wave to
  mitigate parameter misspecifications.}, each obtained with a
$20 \mu s$ annealing process. To adjust for physical temperature of
the hardware, we scaled down the values of $\vc{J}$ by a factor of
$2.5$, which is a crude estimate of the $\beta_{\text{HW}}$ parameter for
this problem. As a result, the model programmed on hardware had all
$\vc{J}$ values within the $[-1,1]$ range as required by the 2000-qubit D-Wave system.

The resulting empirical probabilities of 16 local minima under both
MCMC (green) and QA (blue) sampling are shown in
Fig.~\ref{fig:sampling_flc1}. The abscissa represents the 16 local
minima. The ordinate records the probability of each local
minimum. Red bars show the probabilities of local minima under a
classical Boltzmann distribution. QA empirical probabilities follow the
exact Boltzmann probabilities closely, with a Kullback--Leibler (KL)
divergence of empirical distribution from exact Boltzmann
distribution of $KL(B\|P_{\mathrm{QA}}) = 0.0011$. In contrast, MCMC annealing
substantially over-samples excited states with corresponding
$KL(B\|P_{\mathrm{MCMC}}) = 0.2222$. MCMC chains become trapped in excited
minima during the anneal, and are not able to cross barriers between
states as the temperature decreases.

\begin{figure}[h]
\centering
\includegraphics[width=0.65\textwidth]{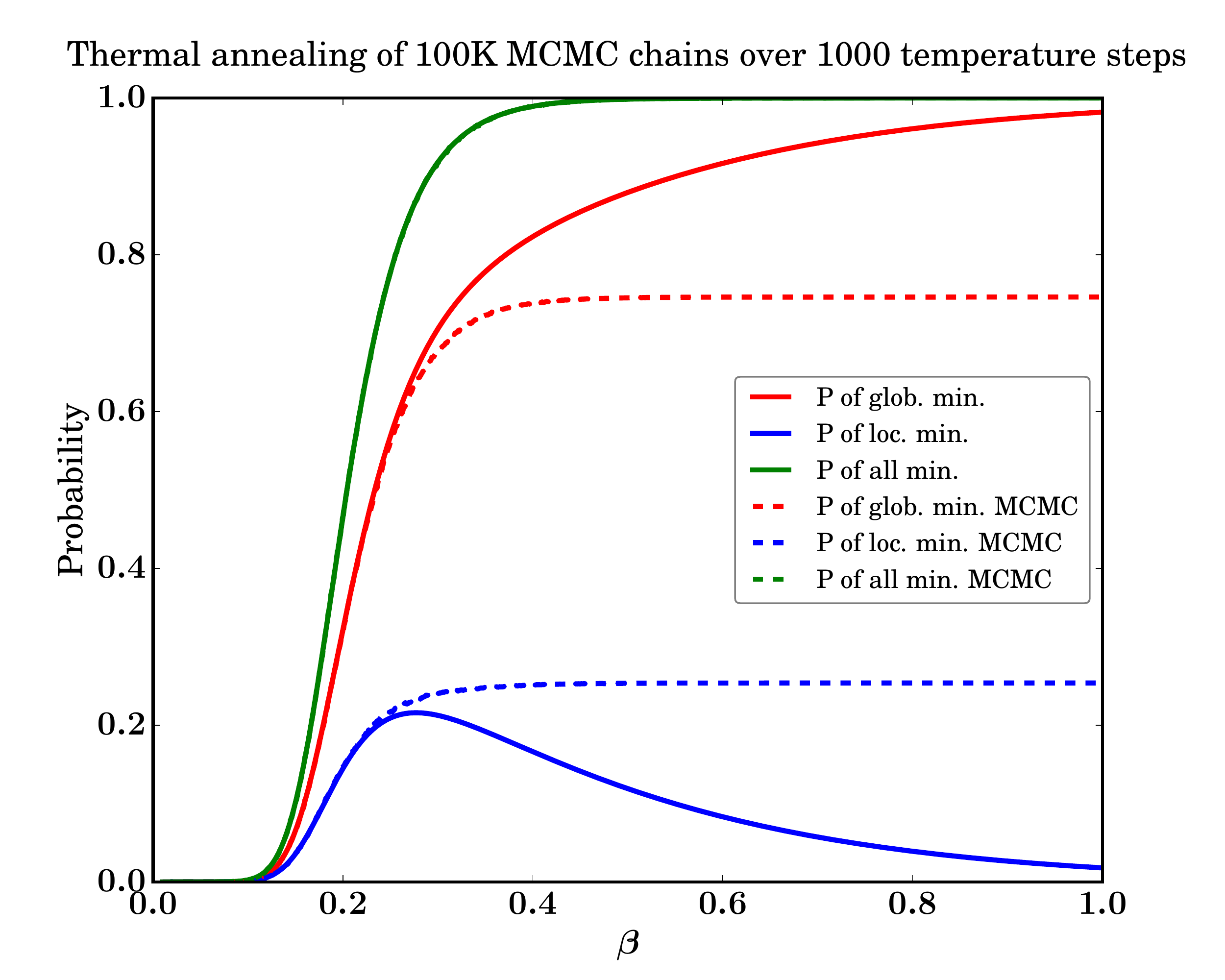}
\caption{Dynamics of the MCMC annealing process on FCL-1.}
\label{fig:dynamics}
\end{figure}

The failure of the MCMC annealing process is shown more in detail in
Fig.~\ref{fig:dynamics}. Here, the abscissa records inverse
temperature, and the ordinate records probability. The solid green,
red, and blue curves represent the exact combined probabilities of all
16 local minima, all 8 ground states, and all 8 excited states
respectively. The dashed lines represent corresponding empirical
probabilities derived from MCMC chains at each temperature
step. Notably, the exact probabilities of excited states change
non-monotonically during the annealing process. At early stages of the
anneal at low $\beta$ values, the probability of excited states
increases as a function of $\beta$ as probability flows from the
entire solution space into the local minima. As $\beta$ increases
further, the dynamics alter. Probability transitions from excited
states to ground states, and the total probability of excited states
decreases as a function of $\beta$. The MCMC process is able to
accurately model probabilities of all states at early stages of the
anneal, but when the energy barriers between states grow sufficiently
large, the process freezes, and the probabilities of local minima do
not change. As a result, MCMC over-samples excited minima.

\begin{figure}[h]
\centering
\begin{subfigure}[b]{0.3\textwidth}
\includegraphics[width=\textwidth]{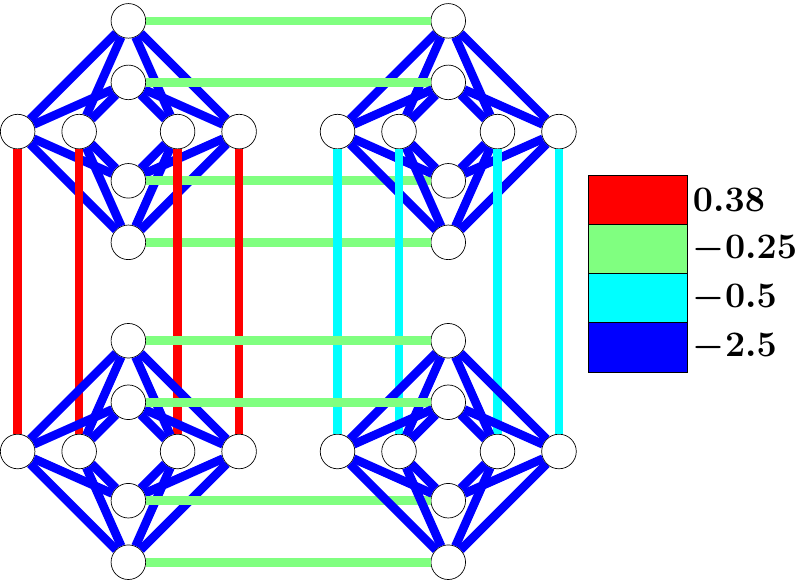}
\caption{Cluster problem with modified inter-cell $J$ couplings (FCL-2).}
\label{fig:flc2}
\end{subfigure}
~
\begin{subfigure}[b]{0.65\textwidth}
\includegraphics[width=\textwidth]{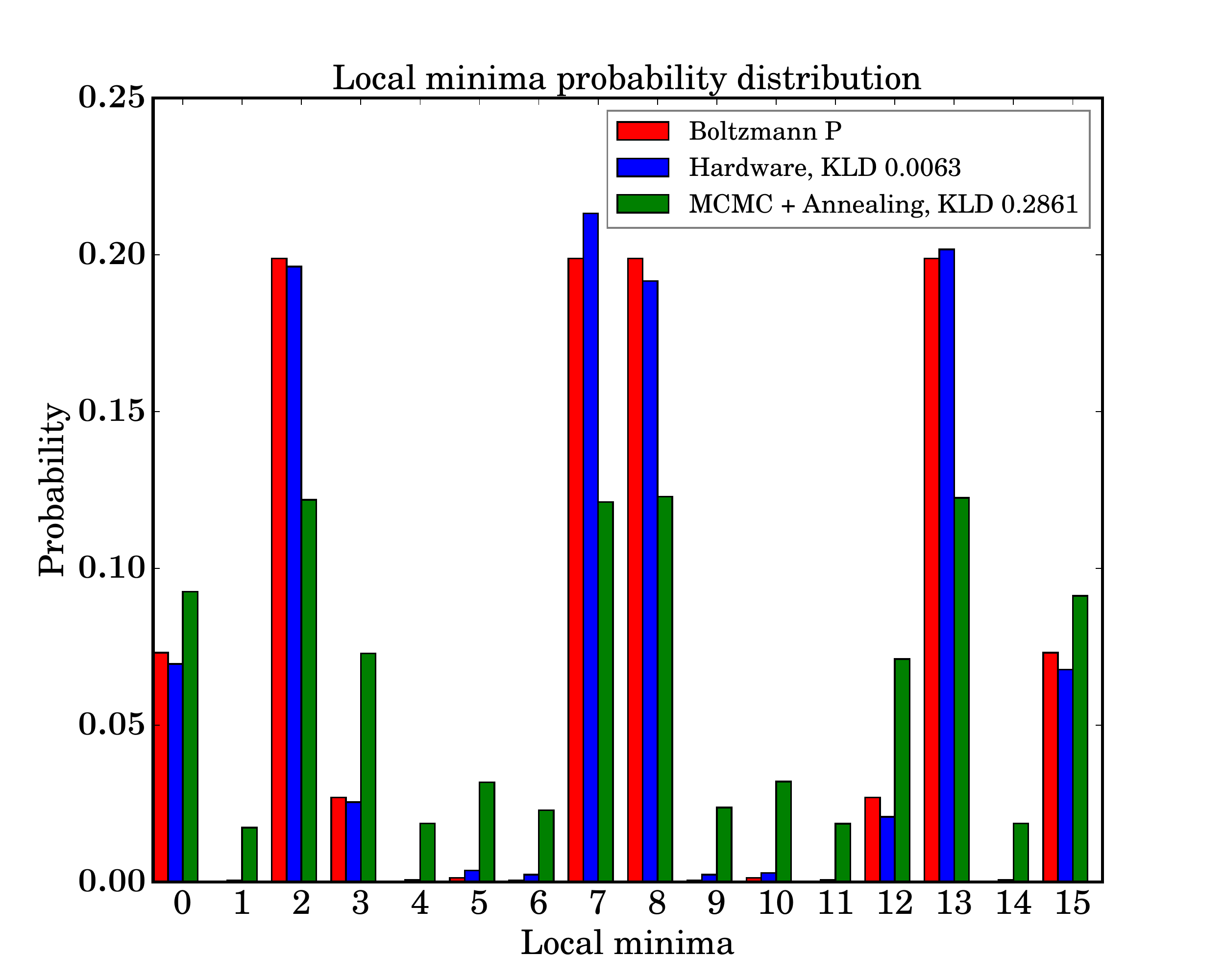}
\caption{Empirical probabilities of local optima obtained from $10^5$
  QA MCMC samples. Red bars represent the exact Boltzmann
  probabilities of local optima, blue and green bars represent
  empirical probabilities derived from QA and MCMC samples
  respectively.}
\label{fig:sampling_flc2}
\end{subfigure}
\caption{FCL-2.}
\end{figure}

It might be argued that a single parameter, $\beta$, can
be adjusted to provide a close match between the QA distribution
and the corresponding Boltzmann distribution, since there are only two
relevant distinct energies within FCL-1. To address this concern, we
modified FCL-1 by breaking symmetry within the inter-cell
frustrated loop connections. The modified problem, FCL-2, is shown on
Fig.~\ref{fig:flc2}.

FCL-2 has the same 16 low-energy local optima, but 4 of these are
ground states, and the remaining 12 excited states have diverse energy
values. We repeated the sampling procedures described above using the
same value of $\beta_{\text{HW}} = 2.5$ to adjust the $J$ values programmed
on hardware. The results are presented in
Fig.~\ref{fig:sampling_flc2}. Again we see that the empirical QA
samples closely follow the exact Boltzmann distribution, with KL
divergence of 0.006, while MCMC annealing continues to over-sample
excited states, only reaching a KL divergence of 0.28.

\begin{figure}[h]
\centering
\begin{subfigure}[b]{0.3\textwidth}
\includegraphics[width=\textwidth]{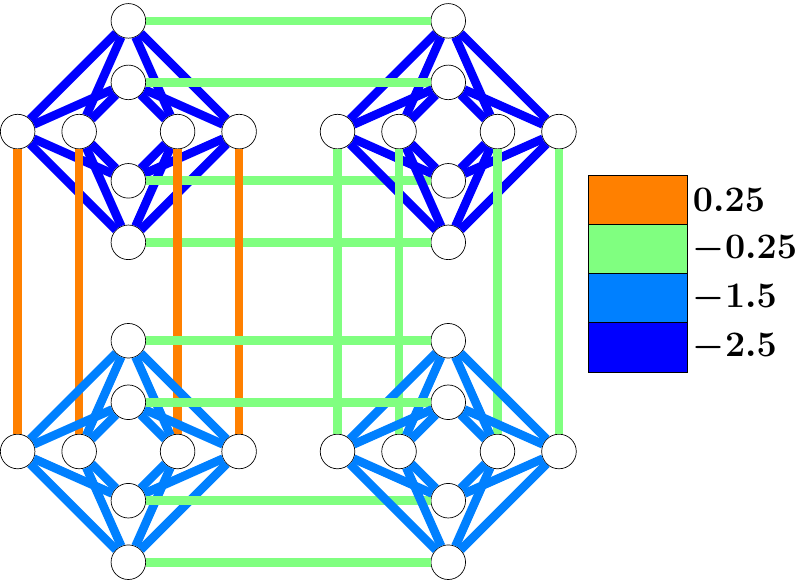}
\caption{Cluster problem with uneven cluster
    strengths (FCL-3). Colour represents weights of the connections.}
\label{fig:flc3}
\end{subfigure}
~
\begin{subfigure}[b]{0.65\textwidth}
\includegraphics[width=\textwidth]{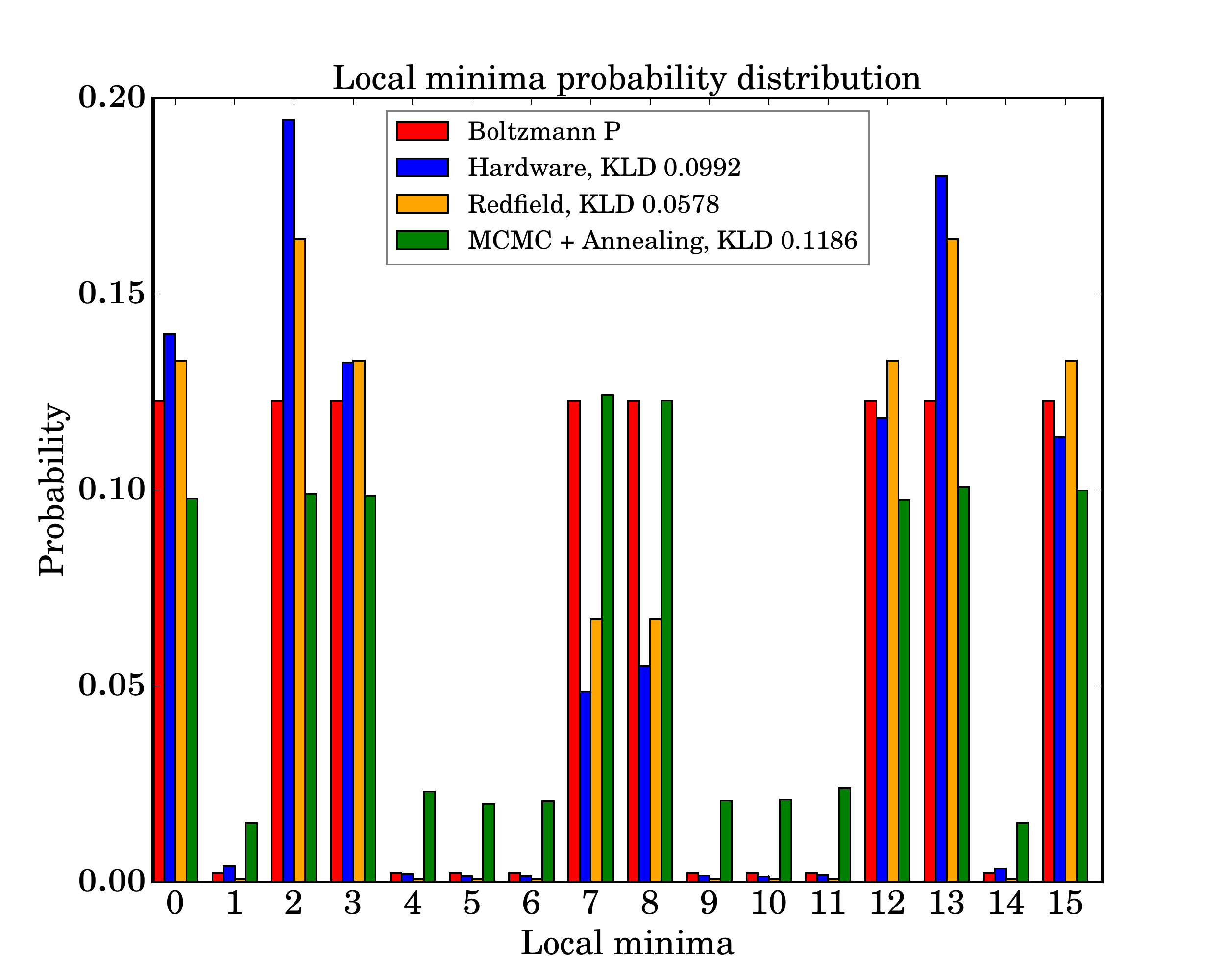}
\caption{The QA distribution deviates substantially from classical
  Boltzmann, but is in a qualitative agreement with the Redfield
  simulation of the quantum dynamics.}
\label{fig:sampling_flc3}
\end{subfigure}
\caption{FCL-3.}
\end{figure}

Thus far, the QA distributions closely approximate the classical
Boltzmann distribution. A little digging into the physics yields the
reason. During quantum annealing, there is a freeze-out analogous to
the classical freeze-out seen in Fig.~\ref{fig:dynamics}. For FCL-1
and FCL-2, the equivalence of all intra-cell interactions means that
quantum effects at the freeze-out point affect all ferromagnetically
connected clusters equally. This freeze-out translates to a simple
energy shift in the classical spectrum, so that the quantum Boltzmann
distribution is very similar to the classical distribution. In general
however, clusters might \emph{not} freeze at the same point. Next, we
consider Ising models where the QA distribution deviates from the
classical Boltzmann. Such models can be obtained by differentiating
among the $J_{\text{intra}}$ couplings. Thus, we consider the FCL-3
problem of Fig.~\ref{fig:flc3}.

The results of the same sampling procedure applied to FCL-3 are
presented in Fig.~\ref{fig:sampling_flc3}. Again, red bars represent
the classical Boltzmann probabilities of energy local minima, and blue
and green bars represent empirical probabilities of local minima
derived from QA and MCMC samples respectively. Now we see that the QA
distribution deviates substantially from classical Boltzmann with a KL
divergence similar to that obtained by a MCMC and anneal procedure
(0.11). Clusters with large (strong) $|J_{\text{intra}}|$ freeze
earlier in the quantum annealing process compared to weak ones
\cite{PhysRevA.92.052323}. Hence, qubits in strong clusters
equilibrate under a quantum Boltzmann distribution at a lower energy
scale than qubits in weak clusters. The result is a distorted
distribution that deviates from the classical Boltzmann. To confirm
this explanation, we applied a classical Redfield simulation of the
quantum dynamics \cite{Amin2009b}. Orange bars in
Fig.~\ref{fig:sampling_flc3} show empirical probabilities of local
minima derived using this simulation agree closely with probabilities
derived from QA samples.

\begin{figure}[h]
\centering
\begin{subfigure}[b]{0.3\textwidth}
\includegraphics[width=\textwidth]{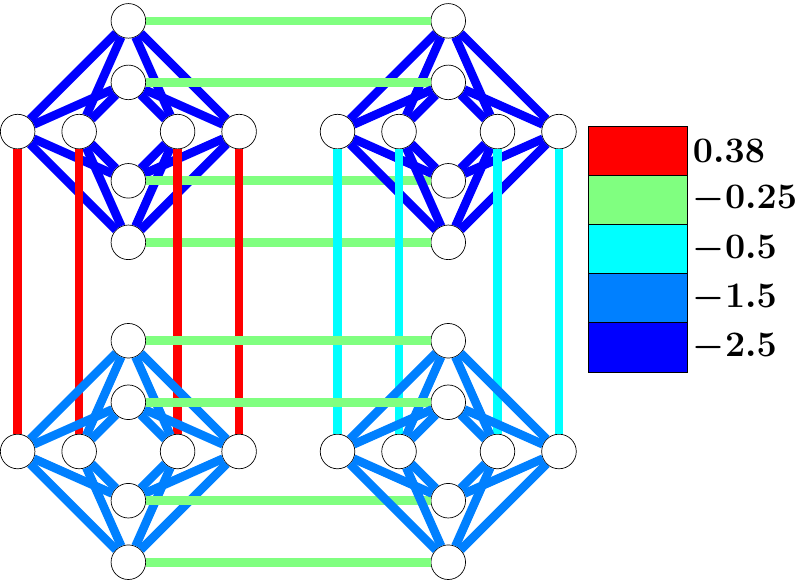}
\caption{Cluster problem with uneven cluster strengths and modified
  inter-cell $J$ couplings. Colour represents weights of the
  connections.}
\label{fig:flc4}
\end{subfigure}
~
\begin{subfigure}[b]{0.65\textwidth}
\includegraphics[width=\textwidth]{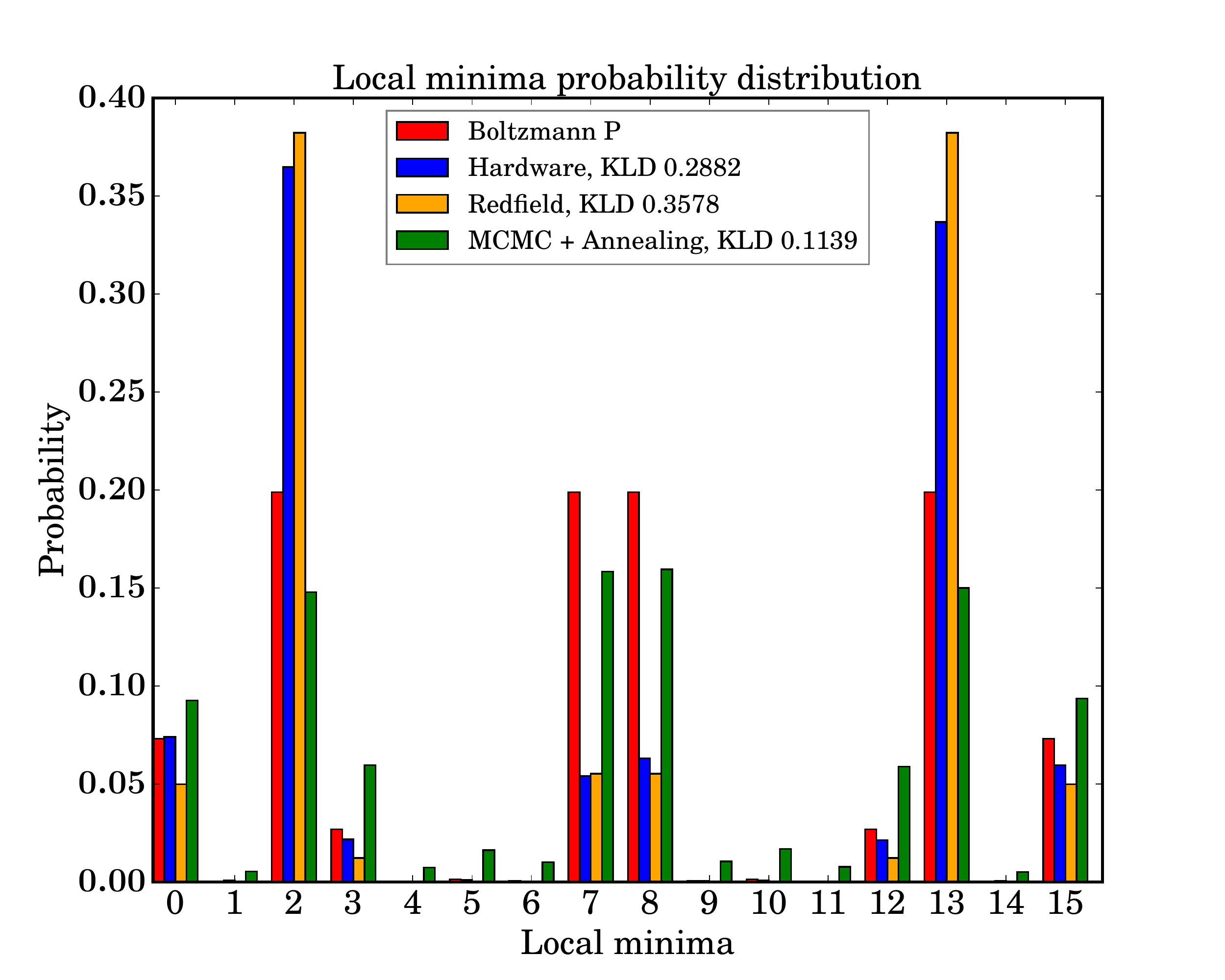}
\caption{Sampling results for FCL-4. The QA distribution deviates
  substantially from classical Boltzmann one.}
\label{fig:sampling_flc4}
\end{subfigure}
\caption{FCL-4.}
\end{figure}

Lastly, we modified cluster strengths for an FCL-2 problem (with a
broken symmetry between excited states) and denoted the resulting
problem FCL-4 (Fig.~\ref{fig:flc4}). The sampling results are shown in
Fig.~\ref{fig:sampling_flc4}. The QA distribution again deviates
significantly from the classical Boltzmann, but agrees closely with the
quantum simulation.

From a machine learning perspective, these asymmetric cluster problems
may appear discouraging, as they suggest that the general QA
distribution has a complicated form that depends on unknown factors,
e.g. freeze-out points for different qubits. In the next
section, however, we show that at least in considered cases it is possible
to adjust (with simple learning rules) hardware parameters to match classical
Boltzmann distributions of interest.

\section{Training Boltzmann Machines Using QA}

\subsection{Fully Visible Boltzmann Machines}
\label{rbm}
A Boltzmann machine defines a probability distribution over
$\pm1$-valued variables $\vc{s}$ as
\begin{equation}
  B(\vc{s} |\vc{\theta}) = \frac{\exp\bigl(-E(\vc{s}|\vc{\theta})
    \bigr)}{Z(\vc{\theta})} \quad \text{with $E(\vc{s}|\vc{\theta}) = \ip{\vc{\theta}}{\vc{\phi}(\vc{s})}$}
  \label{eq:boltzmann}
\end{equation}
where the partition function is
$Z(\vc{\theta}) \equiv \sum_{\vc{s}}\exp\bigl( -E(\vc{s}|\vc{\theta})
\bigr)$. For Chimera-structured BMs the vector of sufficient
statistics is given by
$\vc{\phi}(\vc{s}) = \bigl[ \{s_v\}_{v\in V}, \{s_v
s_{v'}\}_{(v,v')\in E} \bigr]$. Often, hidden variables are introduced
to increase the modeling flexibility of BMs, but we defer the study of
hidden variable models because the likelihood surfaces that result
become multimodal. BMs play an important role in many machine learning
algorithms, and serve as building blocks for undirected generative
models such as deep BMs \cite{SalHinton07}.

In fully visible BMs, the parameters $\vc{\theta}$ are learned from
training data $D=\{\vc{s}^{(i)}\}_{i=1}^{|D|}$ by maximizing the expected
log-likelihood $L(\vc{\theta})$ of $D$:
\begin{gather}
  L(\vc{\theta}) = \mathbb{E}_{P_D(\vc{s})} \bigl( \ln
                   B(\vc{s}| \vc{\theta})  \bigr) = -
                   \bigl\langle \vc{\theta},
                   \mathbb{E}_{P_D(\vc{s})} \bigl( \vc{\phi}(\vc{s})
                   \bigr) \bigr\rangle - \ln Z(\vc{\theta})  \label{log-likelihood-B} \\
  \vc{\nabla} L(\vc{\theta}) =  -\mathbb{E}_{P_D(\vc{s})} \bigl(
                                        \vc{\phi}(\vc{s}) \bigr)
                                        +
                                        \mathbb{E}_{B(\vc{s}|\vc{\theta})}\bigl(
                                        \vc{\phi}(\vc{s}) \bigr) \label{eq:gradients}
\end{gather}
where $P_D(\vc{s}) = \sum_{i=1}^{|D|} [\vc{s}=\vc{s}^{(i)}] / |D|$ is
the training data distribution. Though $L(\vc{\theta})$ is a concave
function (making maximization straightforward in principle), neither
$L$ nor $\vc{\nabla}L$ can be determined exactly for models at large
scale. Thus, training of practically relevant BMs is typically very
difficult.  The dominant approach to training BMs is stochastic
gradient ascent, where approximations to $\vc{\nabla}L$ are used
\cite{Younes98}. MCMC (specifically Gibbs sampling) is used to
estimate
$\mathbb{E}_{B(\vc{s}|\vc{\theta}_t)}\bigl( \vc{\phi}(\vc{s}) \bigr)$
needed for $\vc{\nabla}L(\vc{\theta}_t)$ at parameter setting
$\vc{\theta}_t$, and $\vc{\theta}_t$ is updated (most simply)
according to the estimated gradient as
$\vc{\theta}_{t+1} = \vc{\theta}_t + \eta_t
\vc{\nabla}L(\vc{\theta}_t)$. A variety of methods are available for
the gradient step size $\eta_t$. The efficacy of stochastic gradient
ascent depends on the quality of the gradient estimates, and two
methods are commonly applied to seed the MCMC chains with good
starting configurations. Contrastive Divergence (CD)
\cite{hinton2002training, carreira-perpinan05} initializes the Markov
chains with the data elements themselves since (at least for
well-trained models) these are highly likely states. Persistent
Contrastive Divergence (PCD) \cite{tielemanPcd2008}, improves upon CD
by initializing the Markov chains needed for $\vc{\theta}_t$ with
samples from the previous chain at $\vc{\theta}_{t-1}$. If gradient
steps on $\vc{\theta}$ are small, it is hoped that samples from
$B(\vc{s}|\vc{\theta}_{t-1})$ rapidly equilibrate under
$B(\vc{s}|\vc{\theta}_t)$.

\begin{figure}[h]
\centering
\includegraphics[width=0.65\textwidth]{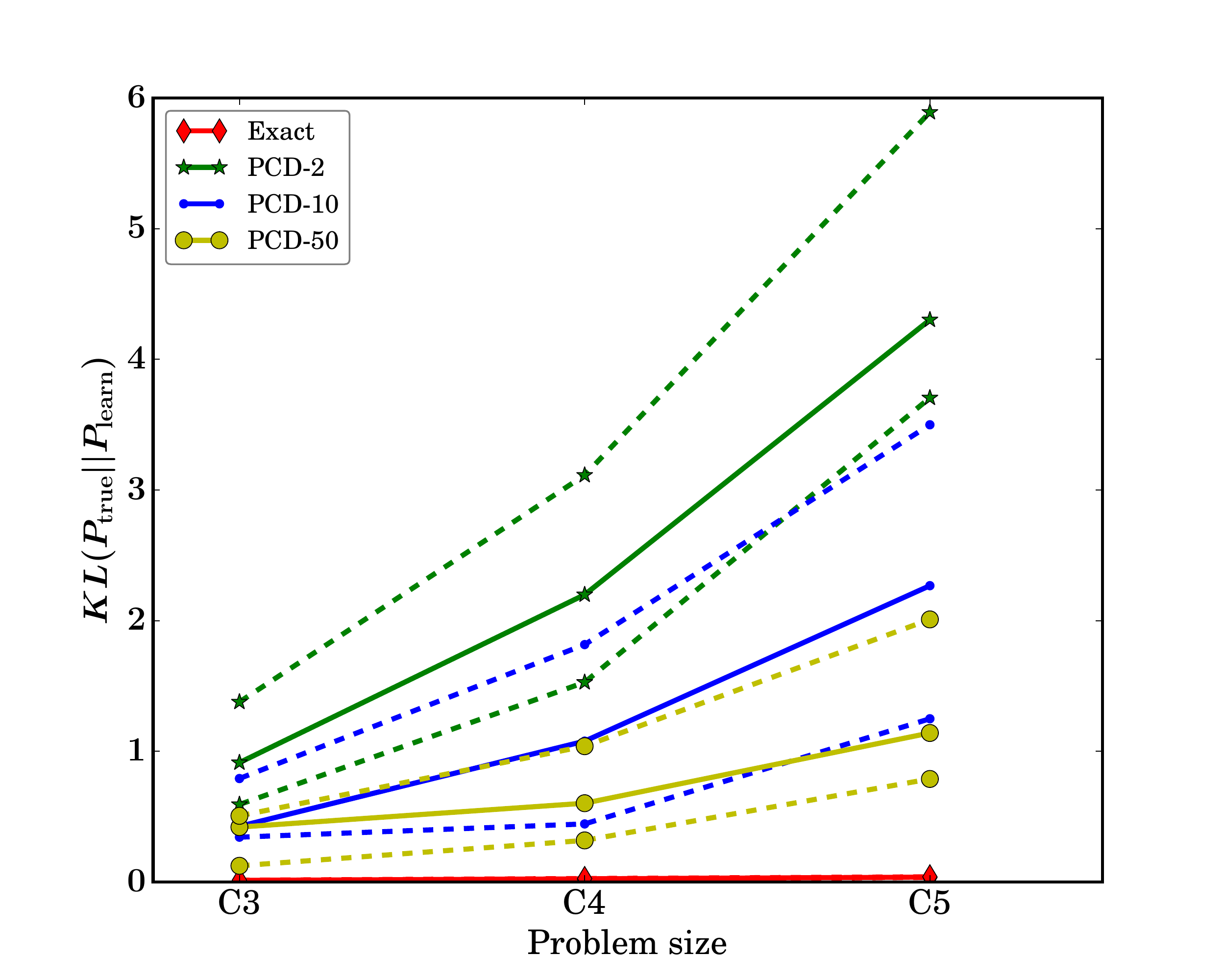}
\caption{Training of random $\theta_{v,v'} = J_{v,v'} = \pm1$
  BMs. Models trained with exact samples minimize the KL divergence, but models
  trained with approximate PCD sampling require progressively more
  Gibbs updates to perform well. Solid lines represent the mean value
  across 20 random instances, and dashed lines represent 25th and 75th
  percentiles.}
\label{fig:perfect_learning}
\end{figure}

The approaches used in CD and PCD to foster rapid equilibration
acutely fail in multimodal probability distributions that have high-energy
barriers. However, even simple problems at modest sizes can show the
effects of poor equilibration under PCD as the problem size grows.  To
demonstrate this, we generated 20 Chimera-structured Ising models with
$\theta_v^{\text{true}}=0$ and $\theta_{v,v'}^{\text{true}}$ randomly
sampled from $\{-1,+1\}$ at sizes $C_3$ (72 variables), $C_4$ (128
variables), and $C_5$ (200 variables). PCD-estimated gradients used
1000 chains with either 2, 10, or 50 blocked Gibbs updates, and all
models were trained for 500 iterations using Nesterov-accelerated
gradients \cite{nesterov1983method}.  The Nesterov method uses
momentum (past gradients), and is more susceptible to noisy gradients
than stochastic gradient descent \cite{Devolder2014}.  The learned
model $\vc{\theta}^{\text{learn}}$ results are presented on
Fig.~\ref{fig:perfect_learning} ($\theta_{v,v'}$ is learned, and
$\theta_v$ is fixed to zero). The abscissa represents problem size,
and the ordinate represents the log-likelihood-ratio
$\ln \bigl[ B(\vc{s}|\vc{\theta}^{\text{true}}) /
B(\vc{s}|\vc{\theta}^{\text{learn}}) \bigr]$ averaged on test
data. Note that this ratio is a sampling-based estimate of
$KL\bigl(B(\vc{s}|\vc{\theta}^{\text{true}}) \|
B(\vc{s}|\vc{\theta}^{\text{learn}}) \bigr)$. The exact model is
recovered when the KL divergence is zero. As expected, models trained
using exact samples achieve a KL divergence close to 0 on all
instances, but PCD requires progressively more Gibbs updates as the
problem size increases.

In subsequent experiments, we explore whether QA may improve upon CD
and PCD by providing MCMC seeds that more accurately sample low-energy
states of $B(\vc{s}|\vc{\theta}_t)$ thus allowing for faster
equilibration and better gradient estimates.

\subsection{Experiments}
\label{BM-experiments}
In training models on QA hardware, it is important to distinguish
$B(\vc{s}|\vc{\theta})$ from the D-Wave QA sampling distribution. By
$P_k(\vc{s}|\vc{\theta})$ we denote the distribution formed by
sampling the QA hardware at parameter $\vc{\theta}/\beta_{\text{HW}}$
followed by $k$ sweeps of blocked Gibbs updates at parameter
$\vc{\theta}$. In particular, $P_0(\vc{s}|\vc{\theta})$ is the raw
hardware distribution at $\vc{\theta}/\beta_{\text{HW}}$, and
$P_\infty(\vc{s}|\vc{\theta}) = B(\vc{s}|\vc{\theta})$. In the
experiments we report, we use $k=50$ blocked Gibbs sweeps.

To test QA for BM learning we train fully visible multimodal
Chimera-structured models. For a variety of problems up to $C_5$ scale
(200 variables),
we specify $\vc{\theta}^{\text{true}}$, draw exact
Boltzmann samples\footnote{We can sample exactly because the treewidth
  of $C_5$ is 20.} from $\vc{\theta}^{\text{true}}$, and try to
recover $\vc{\theta}$ from the samples. We compare the efficacy of CD,
PCD, and QA-seeded MCMC chains. In all CD/PCD/QA cases, each chain is
run for 50 blocked Gibbs updates. To assess the accuracy of the
learned models, we measure the log likelihood on both training and held
out test data, and compare these results to known optimal values.

\begin{figure}[h]
\centering
\includegraphics[width=0.65\textwidth]{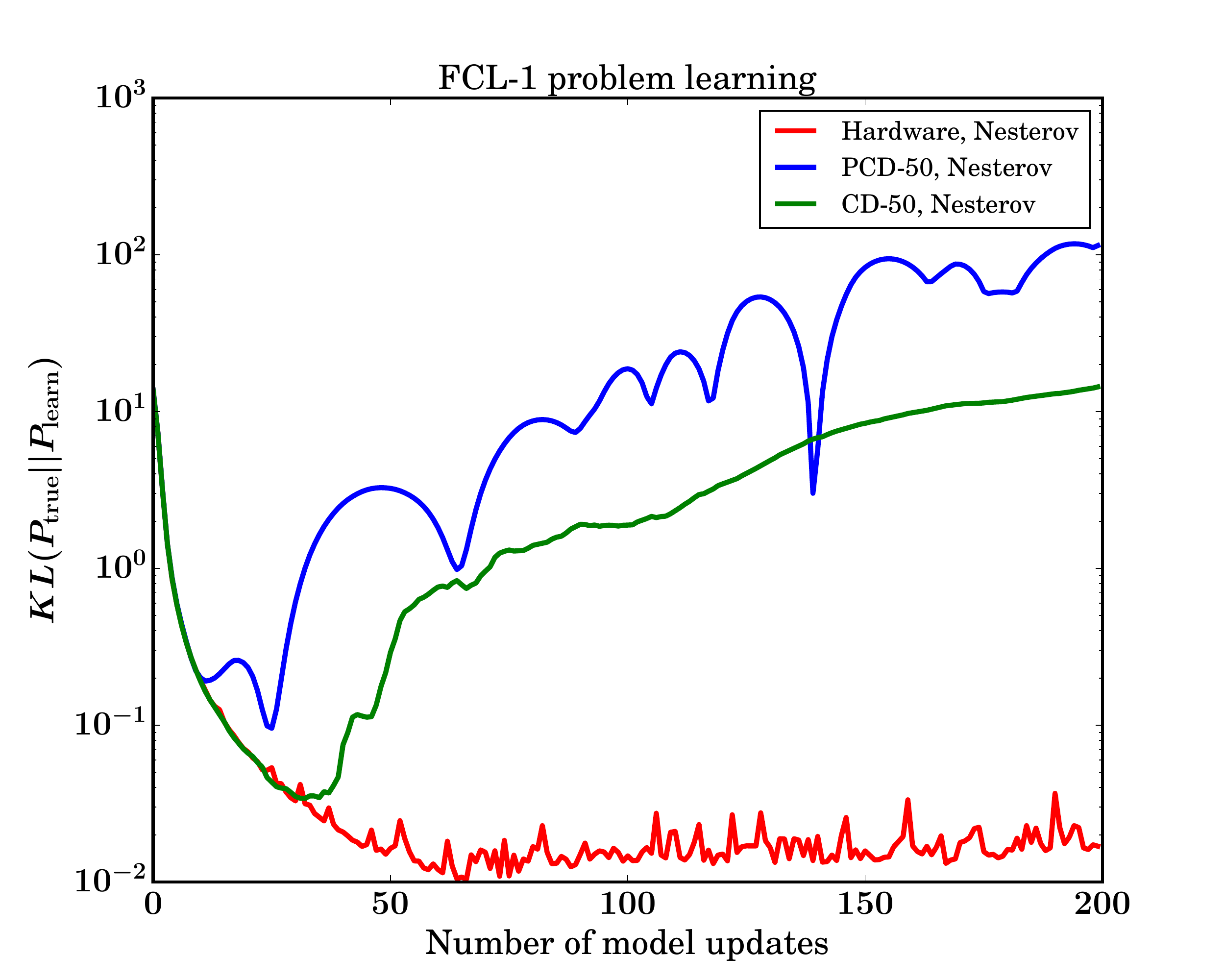}
\caption{Training on FCL-1 using Nesterov-accelerated gradient updates
  with constant step size 0.1 ($\epsilon=0.1$ in the reformulation of \cite{SutskeverMartensDahlHinton_icml2013}).
  Both CD and PCD procedures become
  unstable, but QA-seeded gradients exhibit stable learning.}
\label{fig:training_nesterov}
\end{figure}

For each FCL problem,
we generate a training and a test set of size $5\times 10^5$ using an
exact Boltzmann sampler. All FCL problems have $\theta_v=h_v=0$ and
only $\theta_{v,v'}=J_{v,v'}$ parameters are learned. During training,
gradients are estimated from 1000 Monte Carlo chains seeded with CD,
PCD, or QA initializations. The QA seeds are obtained by calling the
quantum hardware with the standard $20 \mu s$ anneal. In all cases, 50
block Gibbs updates are performed on the seeds. To speed training, we
used Nesterov accelerated gradients. The results for FCL-1 are
presented in Fig.~\ref{fig:training_nesterov}. After about 30
iterations, the CD and PCD procedures collapse, and the corresponding
log likelihoods deteriorate. This occurs when the energy barriers
between local optima in the learned model energy landscape become too
large for the MCMC chains to cross efficiently with 50 Gibbs
updates. As a result, MCMC-based procedures obtain biased gradients
and the CD/PCD models drift away from the optimal region. In contrast,
QA-seeded gradients consistently improve the log-likelihood value for
about 70 updates and stagnate within $10^{-2}$ of $KL=0$.

\begin{figure}[h]
\centering
\includegraphics[width=0.65\textwidth]{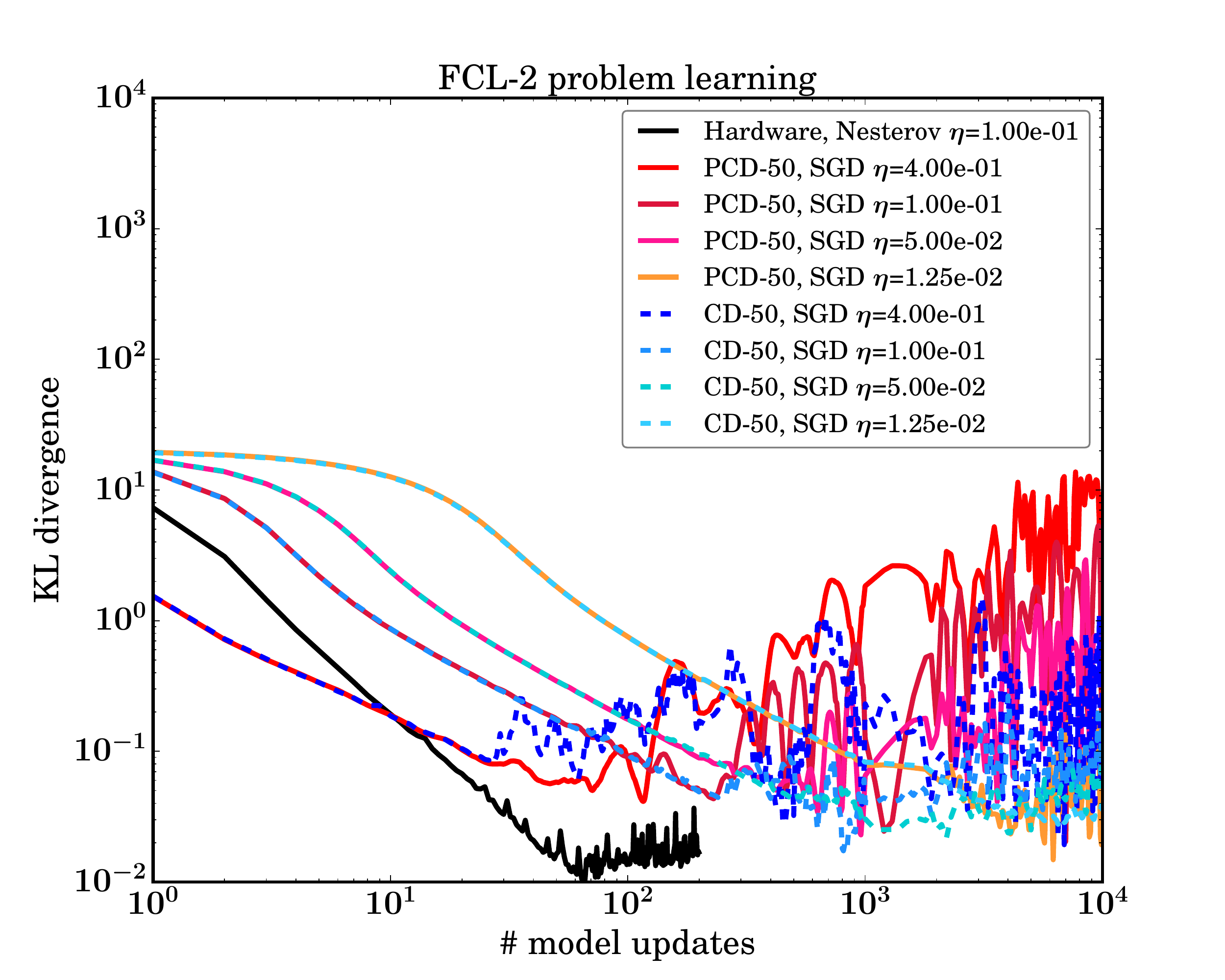}
\caption{Training on FCL-2. QA is trained using Nesterov updates,
  while CD/PCD are trained using standard stochastic gradient descent
  with a fixed learning rate. Decreasing the learning rate for CD
  and PCD improves the stability of the procedures, but increases the
  number of iterations required to reach low values of KL divergence.}
\label{fig:training_sgd}
\end{figure}

The poor performance of CD and PCD is due in part to the choice of the
Nesterov accelerated gradient updates, which, as mentioned earlier, are more
sensitive to noisy gradients than stochastic gradient descent updates.
Interestingly, increasing the number of Gibbs steps (up to $10^6$) does
not help either CD or PCD significantly.
As expected, we found training CD/PCD with simple
stochastic gradient updates to be more effective over a wide range of
iteration-independent learning rates $\eta_t=\eta$. A smaller learning
rate effectively corresponds to a larger number of Gibbs updates at a
larger learning rate, and therefore improves the quality of estimated
gradients, but takes more time. We trained CD/PCD models for 10,000
iterations, and compared to 200 iterations of training using QA with
Nesterov-accelerated gradients. The CD/PCD learning rates were varied
from $\eta=0.4$, where learning rapidly goes unstable, to
$\eta=0.0125$ where learning was impractically slow within 10,000
iterations. The results are shown in Fig.~\ref{fig:training_sgd}. We
found that some of the CD and PCD trained models achieved $KL$ values
similar to that of QA-based learning, but required $10^2$ times as
many model updates.

It is reassuring that QA samples are able to improve upon CD/PCD in
FCL-1 and FCL-2 where the QA distribution closely follows the
classical Boltzmann distribution (see Figs. \ref{fig:sampling_flc1}
and \ref{fig:sampling_flc2}). However, what about training on FCL-3
where QA exhibits strongly non-Boltzmann behavior (see
Fig. \ref{fig:sampling_flc3})? In order for the difference in cluster
strengths to be reflected in the data, we scaled down all
$J_{\text{intra}}$ in FCL-3 by a factor of 3.\footnote{The FCL-3
  $J_{\text{intra}}$ weights are strong enough that there are
  negligibly few broken intracluster bonds, and therefore training
  data generated for FCL-3 and FCL-1 are almost identical.} We train
a BM using QA-seeded gradients and fixed learning rate $\eta=0.1$ on
the resulting problem to learn parameters
$\vc{\theta}^{\text{learn}}$.

\begin{figure}[h]
\centering
\includegraphics[width=0.65\textwidth]{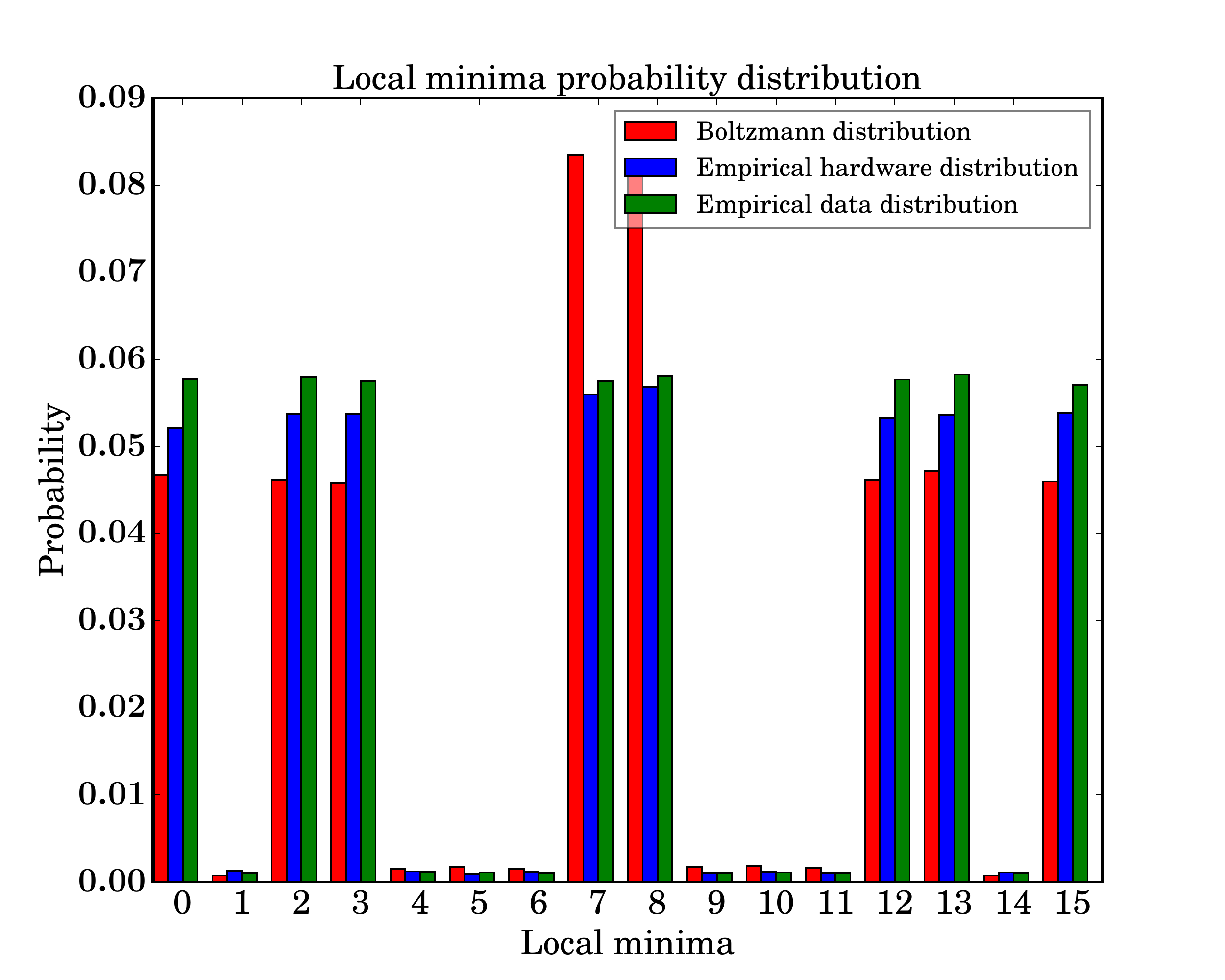}
\caption{Training on scaled-FCL-3 where $J_{\text{intra}}$ parameters
  are scaled down from FCL-3 by a factor of 3. The bars indicate the
  local minimum probabilities derived from the learned model
  $\vc{\theta}^{\text{learn}}$ using a Boltzmann distribution (red),
  and the hardware distribution
  $P_{50}(\vc{s}|\vc{\theta}^{\text{learn}})$ (blue). Green bars are
  the probabilities in the training data.}
\label{fig:flc3_learning}
\end{figure}

To characterize $\vc{\theta}^{\text{learn}}$, we determine the
occupation of local minima under
$B(\vc{s}|\vc{\theta}^{\text{learn}})$ (in red) and
$P_{50}(\vc{s}|\vc{\theta}^{\text{learn}})$ (in blue). In
Fig.~\ref{fig:flc3_learning} green bars represent the local minima
occupation probabilities in the scaled-FCL-3 training data.
The occupation probabilities do not sum to 1 as there is significant
probability of occupying states with broken intracluster bonds.  The
Boltzmann distribution $B(\vc{s}|\vc{\theta}^{\text{learn}})$ fits the
data poorly, but $P_{50}(\vc{s}|\vc{\theta}^{\text{learn}})$ fits the
data well. More detailed examination reveals that
$B(\vc{s}|\vc{\theta}^{\text{learn}})$ over-samples the states that
are under-sampled when the QA hardware is used to sample from FCL-3
(Fig.~\ref{fig:sampling_flc3}). The learning procedure therefore
adjusts the model to compensate for the deviation of QA distribution
from classical Boltzmann. This suggests two important
conclusions. Firstly, the gradients of the loss function
Eq.~\eqref{eq:gradients} used in the training procedure and derived
under the assumption of classical Boltzmann distribution remain useful
in optimizing the model under non-Boltzmann QA distribution. Secondly,
the parameters of the hardware distribution in this case are flexible
enough to closely approximate a classical Boltzmann distribution of
interest.

\section{Assessment of Learned QA Distributions}
\label{alternative_measures}
The results of the previous section suggest that the learned models
$\vc{\theta}^{\text{learn}}$ may not be good fits to training data
under Boltzmann assumptions, but may be when sampling according to
$P_k(\vc{s}|\vc{\theta}^{\text{learn}})$.  Ideally, we would quantify
this by measuring log likelihood on test data, but this is not
directly possible because a closed form expression of the hardware
distribution is unavailable. Instead, we fit a density estimate to
data sampled from $P_k(\vc{s}|\vc{\theta}^{\text{learn}})$, and
evaluate test set log-likelihood using the tractable fit.

\begin{figure}[H]
\centering
\begin{subfigure}[b]{0.9\textwidth}
\includegraphics[width=\textwidth]{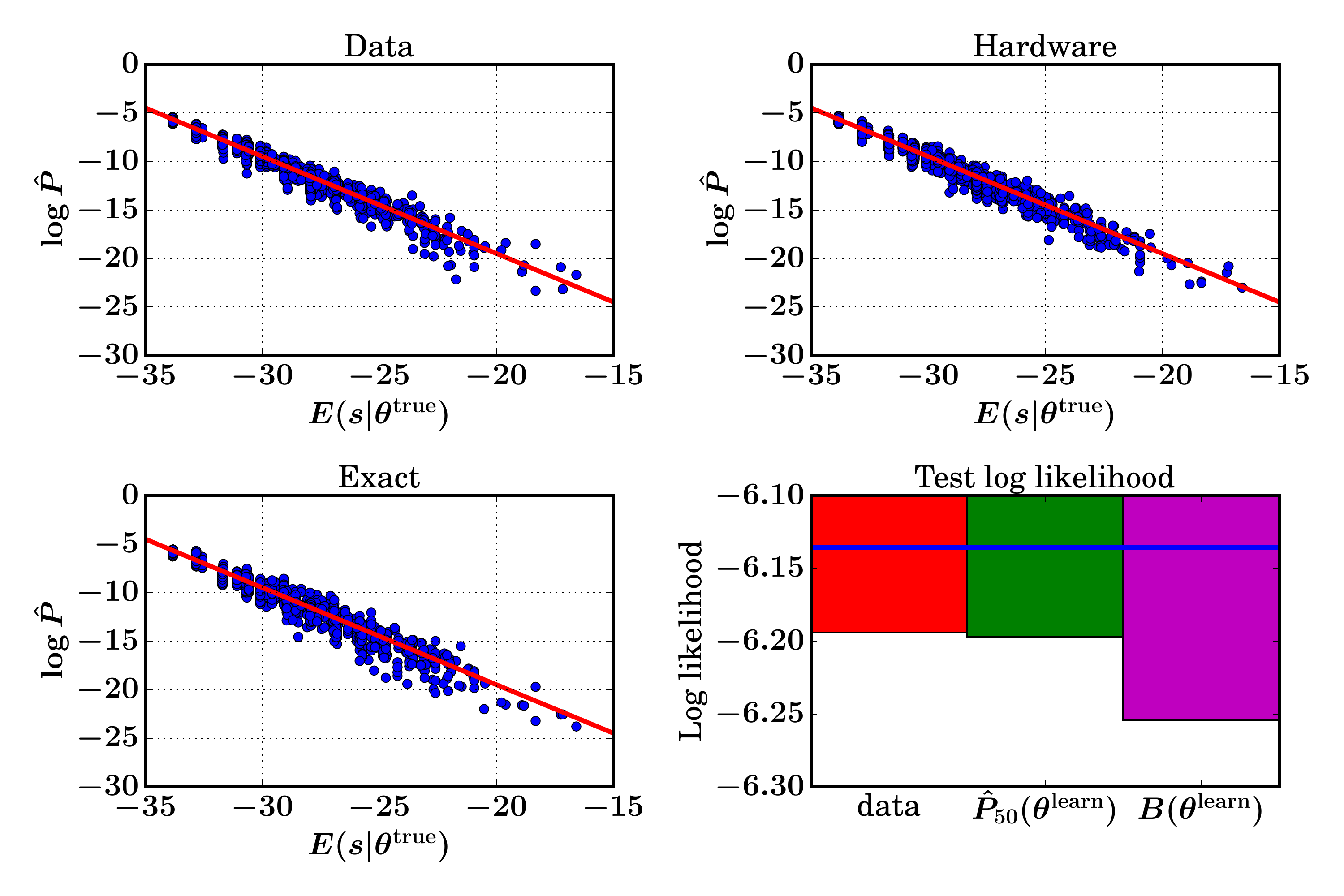}
\caption{NADE estimates on test data.}
\label{fig:nade_data}
\end{subfigure}
~
\begin{subfigure}[b]{0.9\textwidth}
\includegraphics[width=\textwidth]{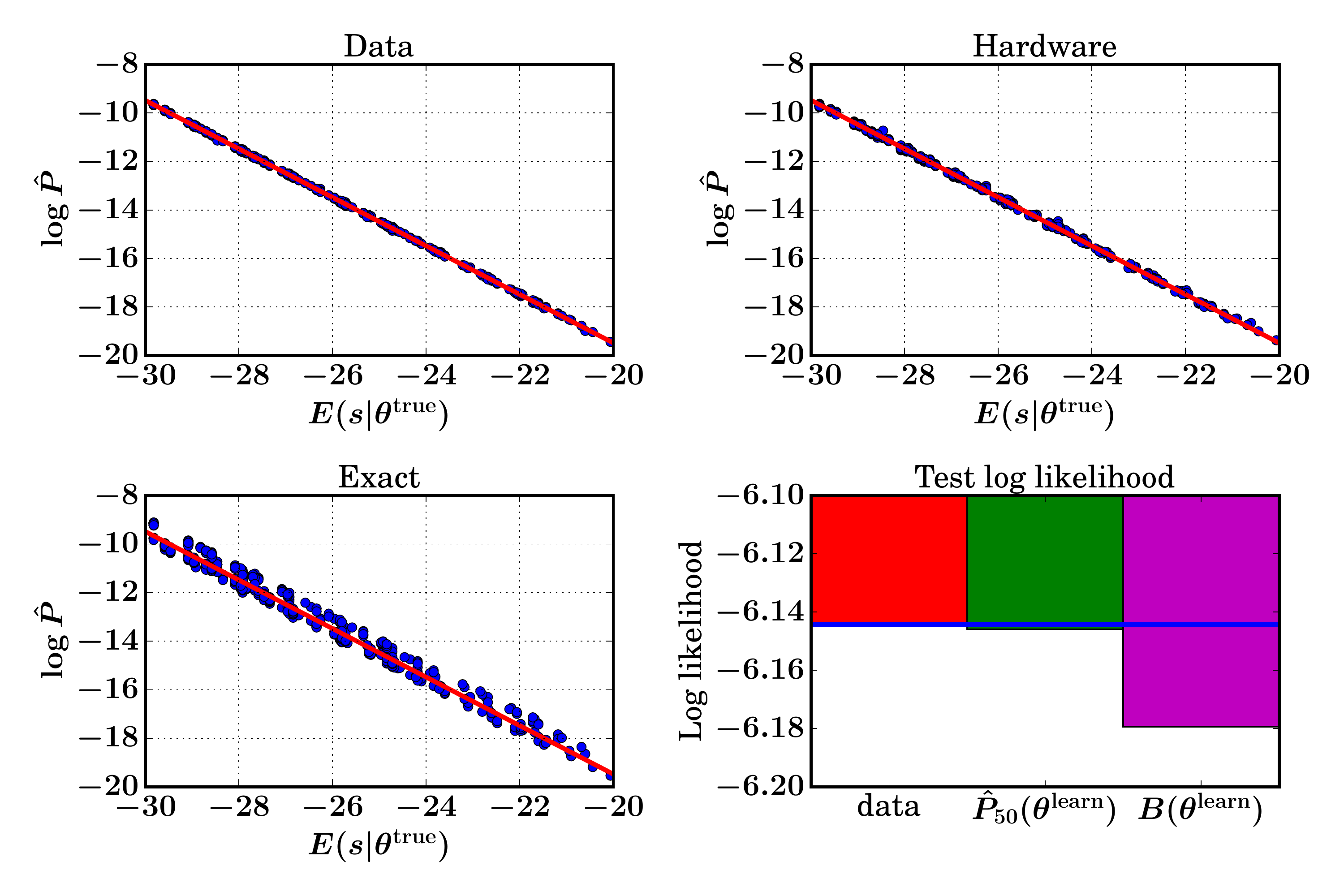}
\caption{Boltzmann estimates on test data.}
\label{fig:orang_data}
\end{subfigure}
\caption{Analytic density estimates.}
\end{figure}

Let $\hat{P}_k(\vc{s}|\vc{\theta})$ represent a tractable fit obtained
from samples of $P_k(\vc{s}|\vc{\theta})$, which approximates
$P_k(\vc{s}|\vc{\theta})$. We require that
$\hat{P}_k(\vc{s}|\vc{\theta})$ can be evaluated for any $\vc{s}$ so
that the log likelihood of test data may be computed. One choice for
$\hat{P}_k(\vc{s}|\vc{\theta})$ is the neural autoregressive density
estimator (NADE) \cite{larochelle11:_neural_autor_distr_estim}. NADE
decomposes the joint distribution into a product of conditional
density estimates, one for each dimension of $\vc{s}$. NADE often
outperforms other density estimators, but it suffers from slow
training and the necessity of hyperparameter tuning. We made some
effort to optimize hyperparameters, but improved values are likely
possible.

Consider again the FCL-3 problem. We denote the FCL-3 parameters by
$\vc{\theta}^{\text{true}}$, and the parameters of the model learned
under QA gradients as $\vc{\theta}^{\text{learn}}$. Let
$B(\vc{s}|\vc{\theta}^{\text{learn}})$ and
$P_{50}(\vc{s}|\vc{\theta}^{\text{learn}})$ represent the Boltzmann
and hardware probability distributions for parameters
$\vc{\theta}^{\text{learn}}$. We compile three data sets each
consisting of $10^4$ samples from
$B(\vc{s}|\vc{\theta}^{\text{true}})$ (data),
$B(\vc{s}|\vc{\theta}^{\text{learn}})$, and
$P_{50}(\vc{s}|\vc{\theta}^{\text{learn}})$. The data sets are further
split into 5000 training and 5000 test points.  To apply NADE to the
datasets, we use an RBM with 200 hidden units with a learning rate
initialized to $0.05$ and decreased over time $t$ as
$1/(1+t/1000)$. The NADE optimization is terminated when the algorithm
sees no performance improvement for 10 consecutive epochs. We validate
the quality of the resultant NADE models by showing scatter plots of
log probability of each test point with respect to its energy (first
three panels of Fig.~\ref{fig:nade_data}). The NADE models are all
roughly Boltzmann with log probability decreasing approximately
linearly with $E$ as expected. In Fig.~\ref{fig:nade_data} we show the
average test set log-likelihood of the NADE models trained on samples
from $B(\vc{s}|\vc{\theta}^{\text{true}})$,
$P_{50}(\vc{s}|\vc{\theta}^{\text{learn}})$ and
$B(\vc{s}|\vc{\theta}^{\text{learn}})$.  For comparison, the horizontal
blue line denotes the likelihood of test data under the true model
$B(\vc{s}|\vc{\theta}^{\text{true}})$. According to NADE, the hardware
model $P_{50}(\vc{s}|\vc{\theta}^{\text{learn}})$ is a better fit to
test data than $B(\vc{s}|\vc{\theta}^{\text{learn}})$.

The NADE algorithm is heuristic and introduces its own error in
estimating the test set log likelihoods, and our hope is that the NADE
error is smaller than the differences in test set log likelihoods. For
models of unknown structure, we have no better alternative than a
blackbox approach like NADE, but on these problems where we know the
training data is Boltzmann distributed we can do better. As all three
distributions should be either Boltzmann or close to Boltzmann, we fit
a Boltzmann distribution to each set of
samples. Fig.~\ref{fig:orang_data} shows analogous results but under a
Boltzmann fit rather than a NADE fit. In this case we see that
$\hat{P}_{50}(\vc{s}|\vc{\theta}^{\text{learn}})$ on test data is an
excellent fit, and almost matches the true test set log
likelihood. Thus, the QA-enabled training procedure learns a very good
data model under the hardware distribution despite the fact that the
hardware distribution is significantly non-Boltzmann. In the rest of
the paper, we assume that $\hat{P}_k(\vc{s}|\vc{\theta})$ is
calculated using Boltzmann estimates.

\begin{figure}[H]
\centering
\begin{subfigure}[b]{0.9\textwidth}
\includegraphics[width=\textwidth]{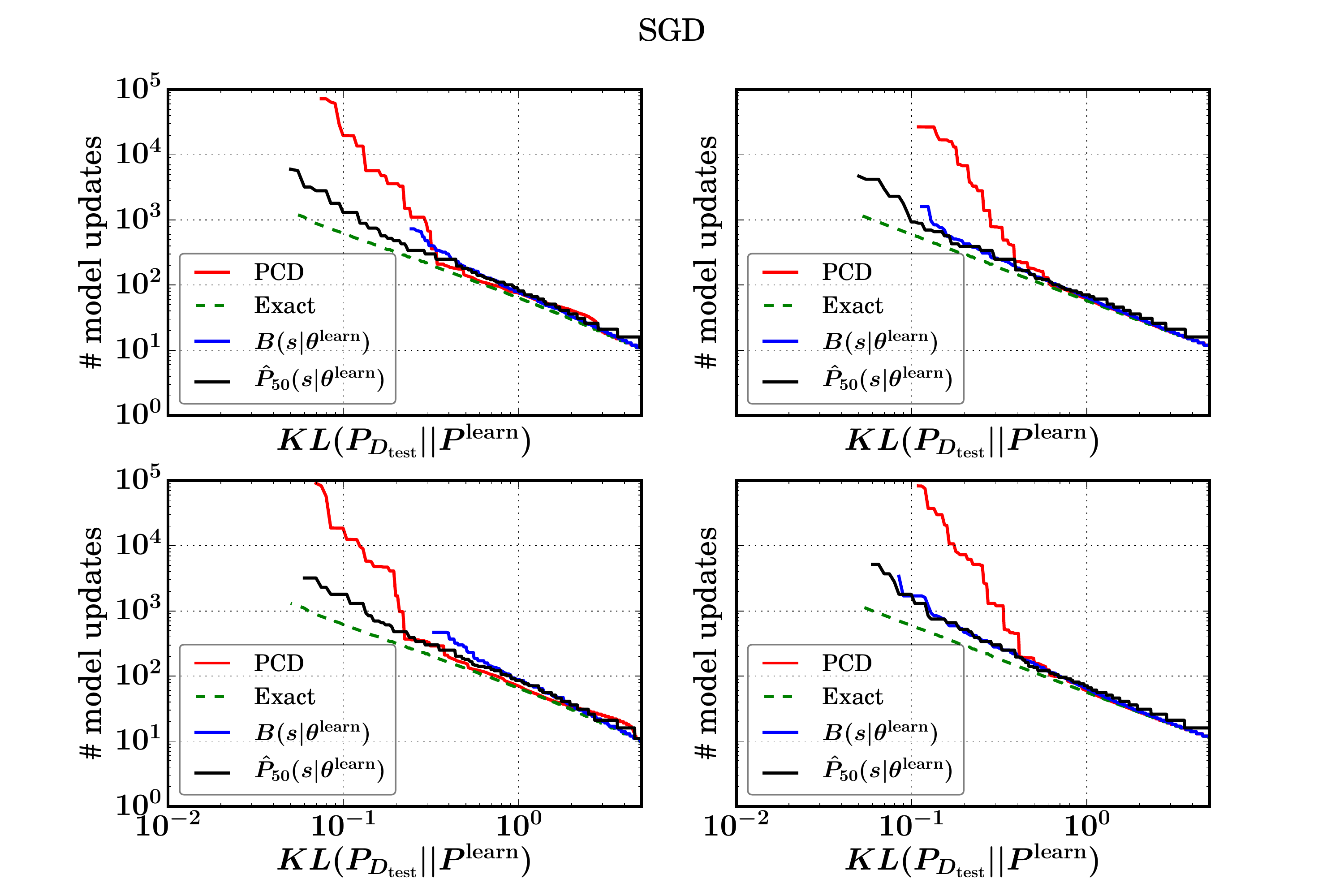}
\caption{Stochastic gradient updates.}
\label{fig:sgd}
\end{subfigure}
~
\begin{subfigure}[b]{0.9\textwidth}
\includegraphics[width=\textwidth]{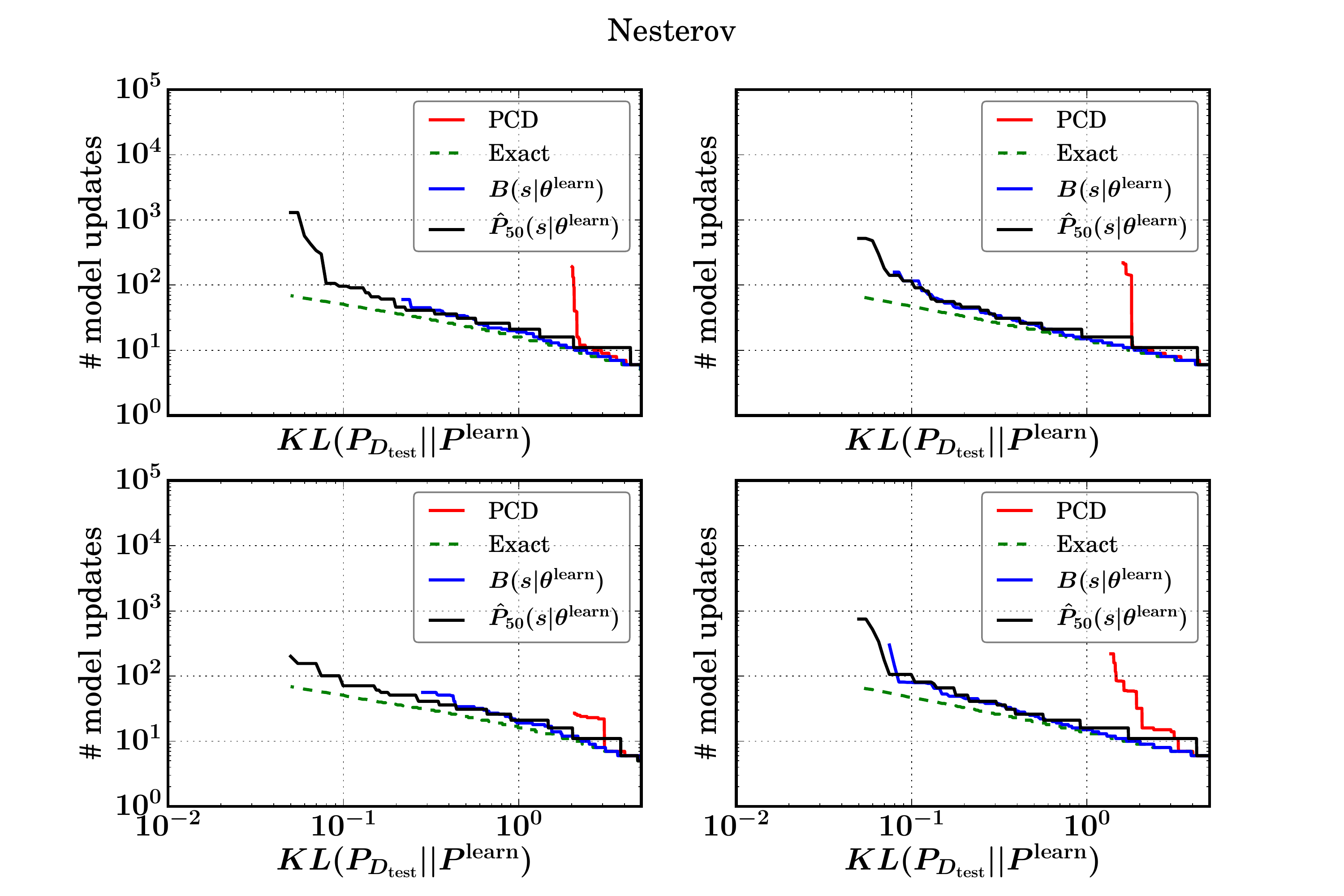}
\caption{Nesterov updates.}
\label{fig:nesterov}
\end{subfigure}
\caption{Learning on four randomly generated $C_5$ frustrated cluster loop
  problems.}
\end{figure}

Lastly, we characterize the relative computational effort of learning
on larger problems. These problems consist of 200 variables arranged
as a $5\times 5$ array of unit cell clusters with
$J_{\text{intra}} = -2.5$, and with inter-cell couplings that are
randomly $J_{\text{inter}}=\pm0.25$. These problems have many local
minima due to the frustrated loops between clusters, and have
high-energy barriers between local minima. We indicate a particular
realization of this model as $\vc{\theta}^{\text{true}}$ and create test and
training states of 500,000 each by sampling from
$B(\vc{s}|\vc{\theta}^{\text{true}})$. Parameters
$\vc{\theta}^{\text{learn}}$ are learned from the training data using
PCD and QA seeded gradients, and approximate KL divergence is measured
using the test data. In all cases, we use 1000 Monte Carlo chains
and apply 50 blocked Gibbs updates. In Figs.~\ref{fig:sgd} and \ref{fig:nesterov} we show the number of
gradient updates required by PCD and QA-seeded gradients to achieve a
specified $KL(P_{\text{true}}|P_{\text{learn}})$ under stochastic
gradient (SGD) and Nesterov updates. We ran PCD at 9 different
learning rates ranging from $\eta = 10^{-1}$ down to
$\eta = 3.9\cdot10^{-4}$, and QA-seeded gradients at learning rates
$\eta = 10^{-1}$, $5\cdot10^{-2}$, and $10^{-2}$. At each $KL$
divergence, we counted the number of gradient updates in the method
requiring the fewest number of updates to attain that $KL$. For
comparison, we also indicate the rate of learning under exact
gradients using a step size of 0.1.

The curves labeled $\hat{P}_{50}(\vc{s}|\vc{\theta}^{\text{learn}})$
and $B(\vc{s}|\vc{\theta}^{\text{learn}})$ are the two variants of
hardware-trained models. Curves that terminate at finite $KL$ values
indicate that no lower $KL$ divergence was found. We see that Nesterov
updates using QA gradients result in the most rapid learning.

\section{Training Quantum Boltzmann Machines Using QA}

We have seen that QA-seeded MCMC can speed training of some classical
BMs. The learning rule we employ,
$\vc{\nabla}L(\vc{\theta}) =
\mathbb{E}_{P_{50}(\vc{s}|\vc{\theta})}\bigl( \vc{\phi}(\vc{s}) \bigr)
- \mathbb{E}_{P_D(\vc{s})}(\vc{\phi} (\vc{s})\bigr)$ (which assumes a
classical Boltzmann sampling distribution), results in models that
adapt to the biases arising from deviations between the QA sampling
distribution and the classical Boltzmann distribution. As a
consequence, $\hat P_k(\vc{s}|\vc{\theta}^{\text{learn}})$ is usually a
better model than $B(\vc{s}|\vc{\theta}^{\text{learn}})$. In light of
this, it is natural to explore a training procedure that avoids
blocked Gibbs postprocessing entirely, namely
$\vc{\nabla}L(\vc{\theta}) =
\mathbb{E}_{P_0(\vc{s}|\vc{\theta})}\bigl( \vc{\phi}(\vc{s}) \bigr) -
\mathbb{E}_{P_D(\vc{s})}(\vc{\phi} (\vc{s})\bigr)$, and evaluate the
generalization of $P_0(\vc{s}|\vc{\theta}^{\text{learn}})$ on test
data.

This may seem a strange learning rule as it is motivated by assuming
the QA sampling distribution is Boltzmann, which it clearly is
not. However, as we show next it can be theoretically motivated.

\subsection{Fully Visible Quantum Boltzmann Machines}

When annealing classically, the dynamics can freeze as the temperature
drops below the size of relevant energy barriers. We provided an
example of this in Fig. \ref{fig:dynamics} for classical annealing on
FCL-1. A similar effect can occur during quantum annealing where
dynamics freeze at time $t$ prior to the end of the quantum anneal at
$t=\tau$. Thus, a more accurate model of QA distribution is described
in \cite{PhysRevA.92.052323} using a transverse Ising Hamiltonian
$\bar{\vc{H}} = \vc{H}(\bar{t})$ for the Hamiltonian of Eq.~\eqref{Ht}
and some $\bar{t} < \tau$.  The density matrix of the distribution
defined by $\vc{\bar H}$ is
\[
 \vc{\rho} = \frac{1}{\bar Z}\exp(-\bar{\vc{H}}),
\]
where the partition function $\bar Z$ is simply the trace of the
matrix $\exp(-\vc{\bar H})$, and the probability of state $\vc{s}$ is
the $\vc{s}^{\text{th}}$ diagonal entry of $\exp(-\vc{\bar H})/{\bar Z}$.
Maximizing the log likelihood $\mathcal L$ of this distribution is
difficult. Instead, \cite{Amin2016} proposes to maximize a lower bound
$\bar {\mathcal L} \le {\mathcal L}$ obtained using the
Golden-Thompson inequality:
\[
 \bar {\mathcal L}(\vc{\theta})  = -
                   \bigl\langle \vc{\theta},
                   \mathbb{E}_{P_D(\vc{s})} \bigl( \vc{\phi}(\vc{s})
                   \bigr) \bigr\rangle - \ln {\bar Z}(\vc{\theta},\Delta).
\]
The gradient of this lower bound can be estimated exactly as in
\eqref{eq:gradients} using the raw QA samples, that is, using
$P_0(\vc{s}|\theta)$:
\begin{equation}
 \vc{\nabla}\bar {\mathcal L}(\vc{\theta}) = -\mathbb{E}_{P_D(\vc{s})} \bigl(
                                        \vc{\phi}(\vc{s}) \bigr)
                                        +
                                        \mathbb{E}_{P_0(\vc{s}|\vc{\theta})}\bigl(
                                        \vc{\phi}(\vc{s}) \bigr). \label{eq:quantumGrad}
\end{equation}

\subsection{Experiments}
\label{sec:anne-learn-rates}

We generated problems as in Section \ref{BM-experiments}.  We focus on
the problem class where QA sampling shows the largest deviation from
classical Boltzmann sampling, namely a $5\times 5$ array of clusters
with randomly assigned cluster strengths from
$J_{\text{intra}}\in\{-1.5,-2.5\}$ as in FCL-4, and where all 4 cycles
are frustrated, and have otherwise random couplings from
$J_{\text{inter}} \in \{-0.5,-0.25,0.38\}$.  Training and test sets
had size $5\times 10^5$ points each, generated by an exact Boltzmann
sampler.  On these problems,
$\hat P_k(\vc{s}|\vc{\theta}^{\text{learn}})$ provides better fits
than $B(\vc{s}|\vc{\theta}^{\text{learn}})$.  As mentioned before, we
used raw hardware samples (postprocessing offered no improvement) and
used $\hat P_0(\vc{s}|\vc{\theta}^{\text{learn}})$ to measure
performance.

\begin{figure}
  \centering
   \includegraphics[width=0.75\textwidth]{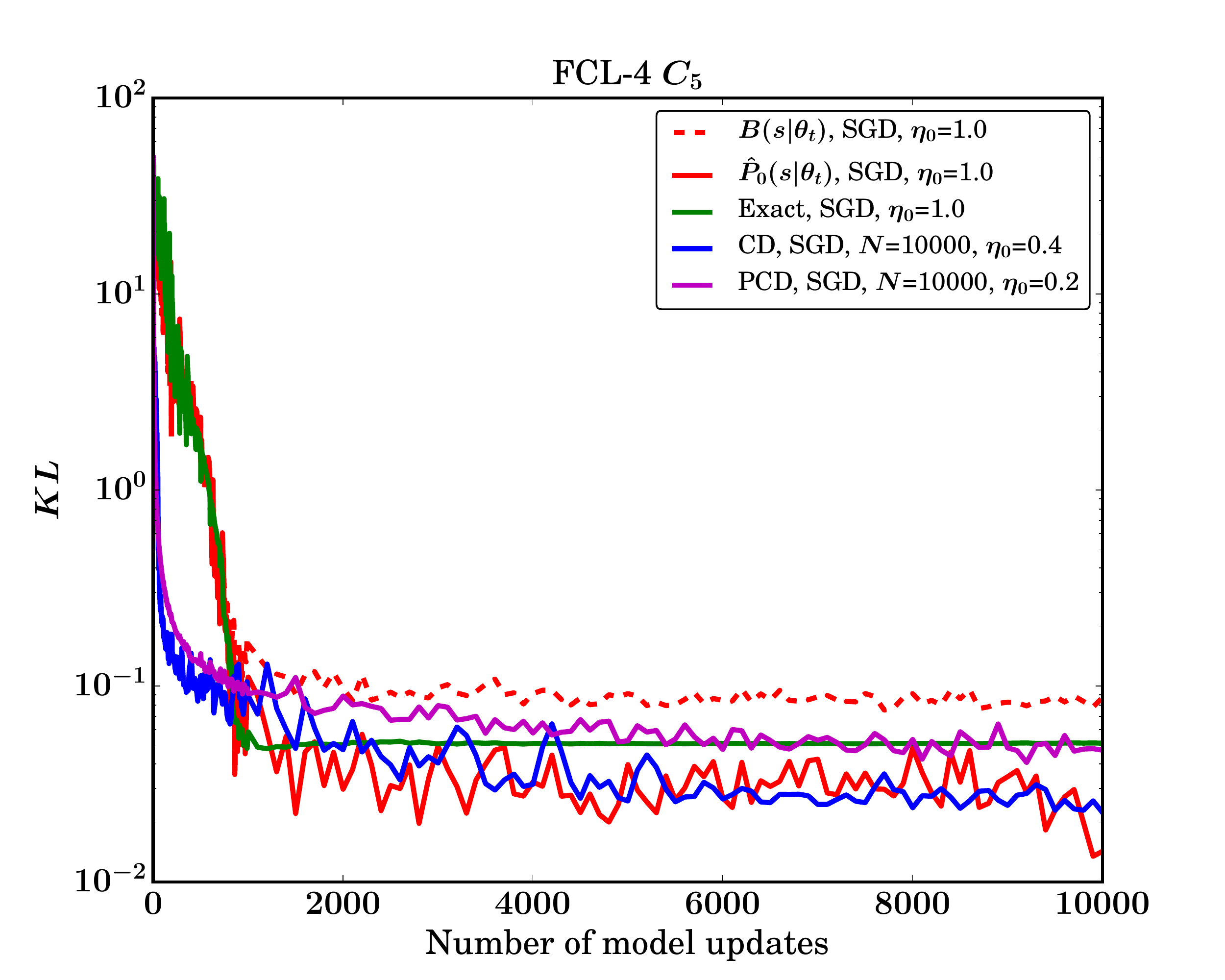}
   \caption{Test set performance under annealed learning
     schedules.} \label{fig:etat}
\end{figure}

We tested annealed learning using gradient step sizes decaying as
$\eta_t = \eta_0/[(t/200) + 1]$.\footnote{The 200 scaling factor was
  determined by cross validation to provide good learning under PCD.}
Both CD and PCD used 10,000 blocked Gibbs updates at each parameter update.
Our findings for $5\times 5$ cluster problems are summarized in
Fig.~\ref{fig:etat}.

These plots show the evolution, over the SGD iterations, of test set KL divergences
$KL\bigl(P_{D_{\mathrm{test}}}(\cdot)\|B(\cdot|\vc{\theta}_t)\bigr)$ for software runs and
$KL\bigl(P_{D_{\mathrm{test}}}(\cdot)\|\hat P_0(\cdot|\vc{\theta}_t)\bigr)$ for QA runs
(the dotted red line is the performance of QA using $B(\vc{s}|\theta)$, for reference).
The $\eta_0$ values shown are the best for each algorithm where
$\eta_0\in \{0.1, 0.2, 0.4, 0.7, 1.0\}$.
For these examples, only CD with 10,000 blocked Gibbs updates was competitive with QA.

\section{Discussion}
\label{discussion}


In this work, we have studied the utility of quantum annealing in
training hard fully visible Boltzmann distributions. We have
empirically characterized the sampling distribution of the D-Wave QA
device on a number of problem classes, and shown that, while the
device is effective at sampling low-energy configurations, the
sampling distribution can differ significantly from classical
Boltzmann.  In spite of this, a learning procedure that updates model
parameters \textit{as if the sampling distribution were Boltzmann}
results in excellent models as long as samples are drawn from the QA
hardware followed by $k$ Gibbs updates. We tested several values of
$k$ and we noticed improvements over CD and PCD. Interestingly, raw QA
samples (i.e., $k=0$) provided similar improvements.  We justify this
by relating learning in classical BMs and quantum BMs as described in
\cite{Amin2016}. We have demonstrated computational benefits over PCD
and CD by measuring the decrease in the number of parameter updates
required for training, and shown benefits under both fixed and
decaying learning rates.

These promising results justify further exploration.  Firstly, the
computational benefits of QA over CD/PCD were demonstrated in
artificial problems constructed to have high-energy barriers between
modes, but which were small enough to yield exact results. We
anticipate that more realistic problems also having large energy
barriers would show similar QA improvement, but this should be
validated.  Secondly, we would like to have further evidence that the
QA model of \cite{Amin2016} or an extension of it can be used to
justify the parameter update rule of Eq. \eqref{eq:quantumGrad} to raw
QA samples. Our motivation is heuristic, and a deeper understanding
might provide more effective learning updates. Thirdly, the sparsity
of connections on current QA hardware limits the expressiveness of
models, and hidden variables are required to model distributions of
practical interest.  Thus, studies similar to this one should
characterize performance for QA-based learning in models with hidden
variables. Lastly, QA hardware is continuously being improved, and new
parameters that control the quantum annealing path (the $\mathcal{A}(t/\tau)$
and $\mathcal{B}(t/\tau)$ functions of Eq.~\eqref{Ht}) have recently been
developed. Learning to exploit these additional controls for
improved training is an important and challenging task.

\bibliographystyle{alpha}
\newcommand{\etalchar}[1]{$^{#1}$}

\end{document}